\documentclass[12pt]{article}
\usepackage{graphicx}
\usepackage[margin=1in]{geometry}
\usepackage{microtype} 

\usepackage[breaklinks, colorlinks=true, urlcolor=blue, citecolor=blue]{hyperref}
\usepackage{natbib}
\usepackage{setspace}
\usepackage{mathptmx}
\usepackage[flushleft]{threeparttablex}
\usepackage{booktabs}
\usepackage{wrapfig}
\usepackage{listings}
\usepackage{amsmath}
\usepackage[hang,flushmargin]{footmisc}
\usepackage[T1]{fontenc}
\usepackage{longtable}
\usepackage{caption}
\usepackage{booktabs}
\usepackage[table]{xcolor}
\usepackage{silence}
\setcitestyle{authoryear,open={(},close={)}}

\title{
    \Large Collaborating with AI Agents:\\ \vspace{-0.5em}
    \Large A Field Experiment on Teamwork, Productivity, and Performance
}
\author{
    \vspace{-1.6em} \small Harang Ju\\ 
    \vspace{-1.6em} \small Johns Hopkins Carey Business School \\ 
    \vspace{-1.6em} \small harang@jhu.edu 
    \and 
    \vspace{-1.6em} \small Sinan Aral\\ 
    \vspace{-1.6em} \small MIT Sloan School of Management \\ 
    \small sinan@mit.edu
}
\date{\normalsize\today}

\begin{document}

\maketitle

\begin{abstract}
\setstretch{1.1}
We examined the mechanisms underlying productivity and performance gains from AI agents using a large-scale experiment on Pairit, a platform we developed to study human-AI collaboration. We randomly assigned 2,234 participants to human-human and human-AI teams that produced 11,024 ads. We evaluated the ads using independent human ratings and a field experiment on \textit{X} which garnered $\sim$5M impressions. We found human-AI teams produced 50\% more ads per worker and higher text quality, while human-human teams produced higher image quality, suggesting a \textit{jagged frontier} of AI agent capability. Human-AI teams also produced more homogeneous outputs (\textit{diversity collapse}) as their ads were more self-similar. The field experiment revealed higher text quality (produced by human-AI teams) increased click-through rates and view time, while higher image quality (produced by human-human teams) improved cost-per-click rates. Two mechanisms explained these effects: Human-AI collaboration was significantly more \textit{task-oriented} and displayed more \textit{delegation}. AI-appropriate work processes, in turn, improved performance in human-AI teams and transcended the jagged frontier: Task-oriented communication produced higher productivity, higher quality ads across all dimensions, and displayed no diversity collapse in their outputs, while human-AI teams with more interpersonal communication were less productive and produced lower quality ads across all dimensions. In contrast, in human-human teams, interpersonal communication improved productivity. Delegation improved text quality in human-AI teams and was positively associated with diversity collapse, creating homogeneous outputs of higher average quality. The results suggest AI agents drive productivity, performance, and output diversity by reshaping teamwork along the jagged frontier of AI capabilities.
\end{abstract}

\clearpage

\section{Introduction}\label{sec:Intro}

Artificial intelligence (AI) tools have garnered attention for their potential to improve productivity and performance \citep{eloundou2024gpts, bick2026adoption}. For example, large language models (LLMs) decreased the average time taken for mid-level professional writing tasks by 40\% and increased output quality by 18\% \citep{noy2023experimental}. For job seekers, AI assistance with resumes increased job hiring by an average of 8\% \citep{wiles2025writing}, and for customer support workers, AI assistance increased productivity by an average of 15\% \citep{brynjolfsson2025genai}. Moreover, productivity gains were greater for lower-skilled workers \citep{noy2023experimental, brynjolfsson2025genai, choi2025ai} and varied across task domains \citep{dellacqua2026navigating}. Evidence from an online labor market suggests there has already been a reduction in demand for freelance knowledge work with the advent of generative pre-trained transformer (GPT) models \citep{hui2024shortterm}. In a meta-analysis of 106 experiments, \cite{vaccaro2024combinations} found that human-AI systems outperformed humans alone on average. 

While studies like \cite{liu2023captions} explore levels of proactivity in AI agents, they typically do so without randomized controlled trials (RCTs) measuring productivity effects. The studies that use RCTs to estimate productivity effects tend to randomize access to LLM chatbots, not AI agents [e.g. \cite{dellacqua2026navigating} and \cite{chen2024large}], which are not typically multimodal, do not include context, do not allow the chatbots to take independent actions or use APIs to call outside of the work environment, and do not provide a collaborative workspace where machines and humans can jointly manipulate output artifacts in real time. These innovations are meaningful because, today, AI agents possess all of these features, but the existing scientific literature studies none of them.

Furthermore, we currently lack fine-grained task level insight into how human-AI collaboration changes work processes and communication patterns and how these changes affect productivity and performance. The vast majority of existing research focuses on the productivity effects of GPT chatbots on individual workers \citep{noy2023experimental, dellacqua2026navigating} or how AI changes people's perceptions, beliefs, and behaviors \citep{tey2024judge, costello2024conspiracy}. However, it is unclear how these interactions evolve in real-time collaborations, especially in environments where AI agents can take autonomous actions, adapt dynamically to human input, and participate in tasks requiring creativity and coordination. Our lack of evidence on \textit{in vivo} human-AI collaboration exists, in part, because off-the-shelf experimental platforms and studies do not provide collaborative workspaces where researchers can precisely record and measure the collaboration itself: \textit{e.g.}, transcripts of messages between machines and humans, logs of edits to output artifacts, and API (application programming interface) calls to outside agents or tools. In contrast, current AI applications, such as \href{https://www.notion.so/product/ai}{Notion AI} and \href{https://www.cursor.com/}{Cursor}, already integrate AI agents into such collaborative workspaces and interfaces.

To address these gaps, we developed Pairit, a novel experimentation platform designed to study human-AI collaboration in real-world, extensible tasks. Pairit introduces several key innovations. It enables real-time collaboration between humans and AI agents, allowing participants to manipulate text, images, and workflows collaboratively in chat-enabled workspaces that mirror existing online AI collaboration work processes. The platform supports randomized pairings of humans and AI (\textit{i.e.}, human-human or human-AI teams) and allows for randomization of prompts and model fine-tuning. Critically, the AI can perform the same actions that humans can perform in the collaborative workspace, except for submitting ads. In the marketing experiment analyzed in this study, these include sending chat messages, writing ad copy, editing ad copy, selecting images, and generating new images using an external call to Dall-E 3. Moreover, Pairit captures every time-stamped keystroke, message, edit, swipe, scroll, selection, API call, and intermediate output, providing a rich dataset that allows for the detailed reconstruction of collaboration workflows. This represents a fundamental departure from the existing literature, which enables RCT-based evaluation of the outputs and productivity implications of human collaboration with LLM-based chatbots and co-pilots, but does not enable randomized experiments analyzing the task-level productivity and work process changes created in human collaborations with fully functioning, multimodal AI agents. 

To study the work process, productivity, and performance implications of human collaboration with multi-modal AI agents, we conducted a large-scale randomized study of human-AI collaboration on advertising design and creation, a task requiring creativity, iteration, and precision. A total of 2,234 participants, representative of the U.S. population, were recruited through Prolific and randomly assigned to human-human or human-AI teams using the Pairit platform. Teams worked collaboratively to create marketing campaigns for a think tank's year-end annual report, including generating and selecting ad images and writing ad copy. This process was fully recorded, resulting in a dataset including 11,024 ads, 182,607 messages, 1,889,559 text edits, 62,119 image edits, and 10,074 AI-generated images, offering an unprecedented level of detail with which to understand work processes, communication, productivity and output quality.

Once the lab portion of the experiment was completed, we conducted a field experiment on the ads produced by human-human and human-AI teams. We obtained human and AI quality ratings of the ads, including the quality of the ad copy and images, and the human- and AI-evaluated likelihood of consumer engagement with the ads (measured by click-through rates). We then ran the ads in a real online display ad ecosystem, generating over 4.9 million impressions on \textit{X}, and evaluated click-through rate, cost-per-click, view-through rate, and view-through duration metrics on the ads using the platform's ads API and DocSend's view metrics, which allow us to record how much of the annual report consumers read, page by page, after clicking through on the ads.

Our analyses examine three performance effects documented in the AI literature but in an agentic collaboration context: productivity and performance \citep{noy2023experimental, brynjolfsson2025genai}, the jagged frontier of AI capabilities across tasks \citep{dellacqua2026navigating, gans2026jagged}, and diversity collapse, or the homogenization of AI-assisted outputs \citep{padmakumar2024doeswritinglanguagemodels, chen2024large, hao2026artificial}. We found evidence of all three effects in our agentic setting: human-AI teams produced 50\% more ads per worker, higher-quality text, and more homogeneous outputs, while human-human teams produced higher-quality images. Field tests revealed that higher text quality (from human-AI teams) improved click-through rates and view time, while higher image quality (from human-human teams) lowered cost-per-click (although these results are marginal and not statistically significant), with these offsetting quality effects across the jagged frontier of AI capabilities explaining similar overall field experiment ad performance across team types.

Most importantly, we found that specific teamwork dynamics and collaboration processes---task-orientation, delegation, and AI recognition---moderate these effects. While our treatment randomizes AI agents at the extensive margin (humans working with or without AI), the data we collected on communication patterns, delegation decisions, and AI recognition capture how people interact with, collaborate with, and delegate work to AI agents. Our analysis demonstrates that although teams can collaborate through open dialogue with agentic partners in a shared workspace, AI partners elicit more directive, less interpersonal communication and more task delegation. Data from the \textit{Pairit} platform show that individuals in human-AI teams sent 56\% more messages than those in human-human teams. Furthermore, human-AI teams sent a 24\% higher share of \textit{content-} and \textit{process}-oriented messages, especially messages containing suggestions, instructions, prioritization, and planning. Conversely, human-human teams sent a 23\% higher share of \textit{social} and \textit{emotional} messages, including messages that expressed rapport building, self-assessment, and concern. We find that the performance payoff of communication style depends on partner type. Interpersonal communication was associated with higher productivity in human-human teams, whereas task-oriented communication was associated with higher quality in human-AI teams. We also tracked work process changes: human-AI teams made 73\% fewer direct edits to the copy and delegated 17\% more work to their AI partners. Higher delegation was associated with higher text quality but not image quality, and with lower output diversity. Recognition of the AI partner was associated with more task-oriented and less interpersonal communication. Thus, collaborating with AI agents alters work processes, and these behavioral adaptations may underlie productivity gains and quality effects across the jagged frontier.

Although prior studies have shown that AI tools can improve productivity and reduce task completion times \citep{noy2023experimental, brynjolfsson2025genai}, AI is often treated as a passive tool rather than as an active collaborator in prior research. As AI agents become integral to modern workflows, researchers are beginning to explore their role as work collaborators, rather than mere tools, emphasizing the importance of trust, transparency, and integration in human-AI partnerships \citep{makarius2020rising, anthony2023, collins2024building}. Our work contributes to this emerging literature by presenting the first task-level randomized experiment measuring the work process, productivity and performance implications of human-AI collaboration with fully functional, multimodal AI agents. The Pairit platform serves as the methodological infrastructure enabling this comparison, and the field experiment provides external validation of quality and performance differences. The core contribution is the empirical analysis of how human-AI collaboration reshapes task-orientation, delegation, communication, and recognition, as well as output quality, performance, and productivity. We hope these observations inform future research as AI agents become integral to the workplace.

\section{Theory}\label{sec:theory}

\subsection{Teamwork in Human-Human and Human-AI Teams}\label{sec:theory:teamwork}

Effective teamwork requires more than individual competence and depends critically on how team members coordinate, communicate, and manage interpersonal relationships \citep{marks2001}. These social dynamics generate collective intelligence that exceeds the sum of individual abilities \citep{woolley2010evidence}. A foundational distinction in this literature separates \textit{taskwork}, the technical work itself, from \textit{teamwork}, the processes that enable individuals to work together effectively \citep{salas1992teamwork}. Teamwork encompasses three types of processes: transition processes where teams plan and set goals, action processes where they coordinate and monitor progress, and interpersonal processes where they manage conflict and build motivation \citep{marks2001}. Teams cycle through these phases as they execute work.

Human teams invest substantial effort in all three. They build shared mental models through ongoing communication about goals, strategies, and progress, enabling them to anticipate each other's actions \citep{mathieu2000influence}. This shared understanding allows teams to negotiate who does what and when. Members also build trust, provide support, and resolve conflict to maintain cohesion, though this consumes time and cognitive resources.

The emergence of AI agents that can act autonomously within collaborative workspaces represents a shift from AI as a tool to AI as a teammate. Unlike chatbot-style interfaces where humans initiate each interaction, agents in our setting have full observability of the shared workspace and can read, write, and edit artifacts just as a human teammate would \citep{wang2024agent}. This shift raises questions about how teamwork dynamics change when one team member is artificial \citep{seeber2020machines}. Human-robot interaction offers relevant insights, though most of this work examines physical tasks with clearly delineated roles. Even so, consistent patterns emerge: in manufacturing, healthcare, and military contexts, people adapt their behavior based on perceived robot capabilities and reliability \citep{hancock2011meta}, and trust calibration proves central to effective collaboration \citep{lee2004trust, parasuraman2010complacency}. Poorly calibrated trust leads to either over-reliance on flawed automation or underutilization of capable systems. These dynamics can enable human-AI combinations to outperform either alone, though the magnitude depends on task characteristics and collaboration design \citep{vaccaro2024combinations, collins2024building}. Knowledge work introduces additional complexity: AI agents can perform the same cognitive tasks as human teammates, creating ambiguity about optimal task allocation. At the organizational level, AI integration affects not just dyadic collaboration but broader structures of coordination and role design \citep{bankins2024multilevel}.

The shift from human to AI teammates may alter collaboration. For example, human teammates bring idiosyncratic knowledge, experiences, and perspectives that create cognitive diversity \citep{hong2004diversity}. This diversity allows teams to outperform homogeneous groups of higher-ability individuals, but only when social dynamics like turn-taking and mutual responsiveness allow diverse perspectives to surface \citep{woolley2010evidence}. These same social dynamics carry expectations of reciprocity, fairness, and recognition. AI teammates, by contrast, bring consistent availability and (within their capabilities) reliable execution, but without the diversity that emerges from different life experiences or the social fabric that binds human teams. We examine how these differences manifest in communication patterns and delegation behavior, and how recognition of partner identity moderates these dynamics.

\subsection{Task-Oriented vs Interpersonal Communication in Human-AI Teams}\label{sec:theory:communication}

Human-human collaboration requires social maintenance, including building rapport, providing emotional support, and repairing conflicts when they arise \citep{marks2001, hoegl2001}. These interpersonal processes are not merely overhead; they build the trust and psychological safety that enable effective coordination \citep{edmondson1999psychological}. Psychological safety, a shared belief that the team is safe for interpersonal risk taking, reduces the vulnerability that prevents team members from admitting errors or proposing untested ideas. Yet these processes also consume time and cognitive resources that could otherwise be devoted to task execution.

AI partners require none of this social maintenance. They do not take offense, need encouragement, or require relationship repair after disagreements. This absence of social needs may shift communication patterns and change which communication styles relate to productivity and performance in human-AI teams. When AI agents can take actions on shared work, humans may shift from direct task execution to directing work through communication, adopting a more instrumental stance than they would with human partners \citep{tey2024judge}. We therefore examine whether \textit{communication with AI teammates becomes more task-oriented and less interpersonal}.

Communication patterns may also affect performance outcomes, not just differ across conditions. Team communication predicts performance, though the strength of this relationship depends on task characteristics and team composition \citep{lepine2008teamwork}. One reason is that interpersonal communication builds trust and enables implicit coordination, where members anticipate each other's needs without explicit discussion \citep{rico2008team}. Yet these benefits come at a cost. Attention is a limited resource \citep{kahneman1973attention}, and social maintenance consumes cognitive capacity that could otherwise support task execution. For tasks with clear objectives where implicit coordination matters less than explicit direction, redirecting attention from relationship management to task completion may improve productivity.

Task-oriented communication may prove more effective with AI than with humans because AI teammates do not require negotiation or consensus-building. Rapport-building improves coordination in human-human teams, but interpersonal communication may provide fewer benefits with AI partners who require no social maintenance. We examine here whether \textit{the performance payoff of communication styles depends on partner type, with interpersonal communication supporting performance in human-human teams and task-oriented communication proving more effective in human-AI teams}.

\subsection{Delegation in Human-AI Teams}\label{sec:theory:delegation}

When working with capable partners, individuals face choices about task allocation, deciding what to do themselves versus what to delegate. In human teams, social considerations shape these choices. Norms of fairness discourage assigning disproportionate workloads, concerns about overburdening partners limit requests, and expectations of reciprocity create implicit debts. Economic theory identifies a related constraint. Team production creates moral hazard because individual contributions are difficult to observe \citep{holmstrom1982moral}. Delegation in human teams thus involves interpersonal negotiation and trust, not just task optimization.

AI teammates may alter these dynamics. Without relationships and social standing to maintain and protect, delegation to AI carries fewer interpersonal costs. Prior work shows people adjust their behavior with AI. For example, people respond differently to AI and human recommendations, exhibiting algorithm aversion in some contexts and appreciation in others \citep{dietvorst2015algorithm, logg2019algorithm}, and they reduce effort when AI predictions are confident \citep{agarwal2025sufficient, fugener2022cognitive}. Yet most of this work examines AI as advisor or predictor rather than as a collaborator taking actions on shared work. We examine delegation in settings where AI can take actions directly, editing text, selecting images, and communicating with human partners. Without the social costs of delegating to a peer, humans may delegate more readily to an AI agent. This shift toward increased delegation represents an observable change in how collaborators balance execution and direction \citep{morgeson2010team, seeber2020machines}. We therefore examine whether \textit{humans delegate more work to AI collaborators than to human collaborators}.

Beyond shifting how work is accomplished, delegation may affect output quality. Delegation effectiveness depends on the delegate's competence \citep{leana1986predictors}. This principle extends to AI, which exhibits a \textit{jagged frontier} of capabilities, where AI assistance improves performance for some tasks but worsens it for others \citep{dellacqua2026navigating}. We examine delegation effects across this frontier. Large language models are trained extensively on text and instruction-following, suggesting delegation may benefit text quality. Visual selection requires predicting audience response, a task where AI capabilities are less established, so delegation effects on image quality may be weaker or absent. Human-human teams can resolve disagreements through peer discussion to reach mutually acceptable visual choices, whereas human-AI teams may struggle to replicate this collaborative evaluation process. Consequently, humans might place less trust in AI's image selections, which could lead to asymmetric delegation outcomes across media formats.

Delegation also has downstream consequences for output characteristics. More delegation means more of the final output reflects the partner's contributions. When that partner is AI, delegation amplifies both the benefits of AI collaboration \citep{noy2023experimental, brynjolfsson2025genai} and its costs, including homogeneity \citep{padmakumar2024doeswritinglanguagemodels} and limitations outside AI capabilities \citep{dellacqua2026navigating}. We therefore examine whether delegation is associated with improved output quality and whether these benefits vary with AI capability. Rather than reflecting wrong or suboptimal delegation decisions, delegation outcomes may represent task-capability matching along the jagged frontier, where delegation is beneficial specifically for tasks aligned with the AI's particular strengths.

\subsection{AI Partner Recognition}\label{sec:theory:recognition}

The task-oriented communication and delegation patterns that we observe in human-AI collaboration may not require that participants explicitly recognize their partner as AI, though recognition may amplify these adaptation behaviors. Yet people automatically apply social rules to computers \citep{nass2000machines, reeves1996media}, defaulting to politeness, turn-taking, and concerns about fairness even when interacting with machines. Social-default investment may yield little benefit with AI partners in this setting. Recognition may enable collaborators to override these defaults and match their behavior to the actual demands of the interaction.

Perceived partner identity affects trust, communication, and delegation patterns \citep{zhang2023teammate, glikson2020human}, and appropriate reliance on automation depends on accurate assessment of a partner's capabilities \citep{lee2004trust}. Human-appropriate collaboration strategies include building consensus before acting, softening feedback to preserve relationships, and investing in rapport that enables future reciprocity. These strategies make sense with human partners who have feelings, memories, and participate in ongoing relationships, but they also consume effort without benefit when applied to AI. Participants who recognize their AI partner can instead adopt task-oriented communication (Section \ref{sec:theory:communication}) and delegate more readily (Section \ref{sec:theory:delegation}), iterating quickly without social cost and focusing on task outcomes rather than relationship maintenance. Those who mistakenly believe their AI partner is human may achieve lower performance not because AI collaboration is inherently limited, but because they apply social strategies that fit human partners better than AI partners, rather than adapting to the AI's task capabilities.

Recognition may therefore improve performance by enabling behavioral adaptation. In our design, participants are not told whether their partner is human or AI until after the task, so we can measure the effects of actual and perceived partner identity. Some participants seek conversational cues to infer identity during the session, which motivates measuring belief post-task and treating recognition carefully (see Appendix~\ref{app:interaction_dynamics} for a detailed analysis of identity inference). We therefore examine whether \textit{recognition of an AI partner acts as an amplifier that strengthens task-oriented communication and delegation, and how such adaptation relates to output quality}.

\section{Methodology}\label{sec:Method}

Our study was preregistered and approved by the MIT Committee on the Use of Humans as Experimental Subjects.\footnote{See \href{https://osf.io/jfzha}{osf.io/jfzha} and \href{https://osf.io/95dhu}{osf.io/95dhu}.}

\subsection{The Pairit platform}\label{sec:methods:platform}

\begin{figure}[!ht]
    \centering
    \includegraphics[width=0.9\linewidth]{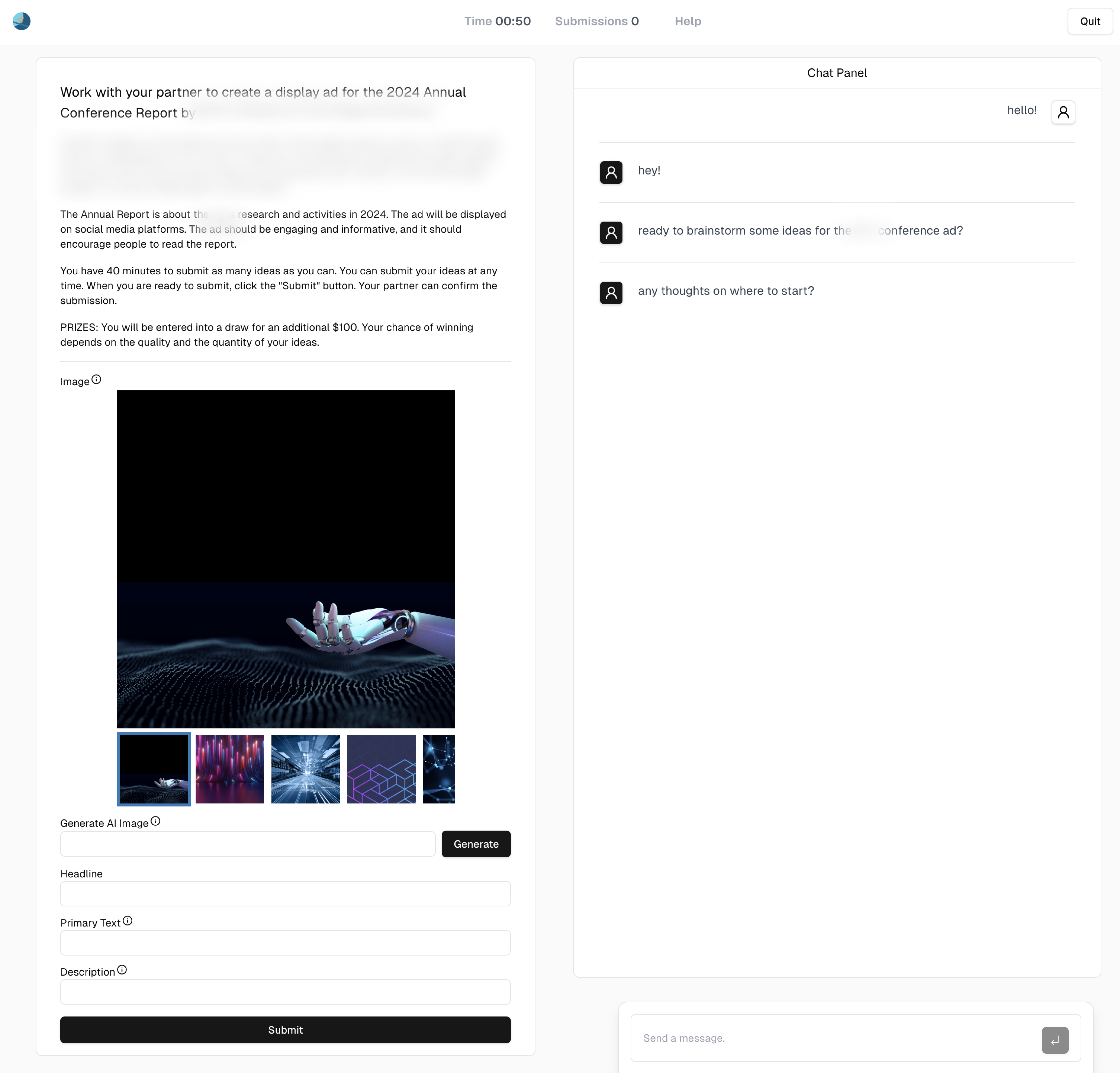}
    \caption{\small The Pairit platform. On the left is the task panel, and on the right is the chat panel. In the human-human condition, chat messages and edits on the task panel (including text edits, image selections, and AI image generations) are synchronized in real-time. In the human-AI condition, the participant chats with an AI agent with full context of the user interface (UI; see Section~\ref{sec:methods:ai:context}), and the AI can edit text, select images, and generate AI images.}
    \label{fig:pairit}
\end{figure}

Once a participant enters the \textit{Pairit} platform (Figure~\ref{fig:pairit}), they are randomized into a queue for either the human-human or the human-AI condition. In the human-human condition, the participant collaborates with another human participant. In the human-AI condition, the participant collaborates with an AI agent. Pairit's left panel is the task panel in which participants can participate in the task work process (in the context of this experiment to create, edit, and submit ads). The platform includes a carousel of seven pre-selected stock images provided by the think tank that the participants and the AI agent can choose from, and the participants (as well as the AI) can also access the Dall-E 3 application programming interface (API) to generate new images. The ad copy creation and editing panel includes a headline, primary text, and a description. All edits, including image selection, image generation, and ad copy changes, are synchronized in real-time across participants. The chat panel displays a ``Typing...'' indicator when the partner is composing a message. When participants submit an ad, the canvas clears and resets for the next submission. On the right is the chat panel in which participants can chat in real time with either another human participant or an AI agent. To the best of our knowledge, Pairit is the first platform to facilitate real-time collaboration between human-human or human-AI teams with team- and model-level randomization and real-time chat and editing on synchronized text and image interfaces in the context of multimodal human-AI collaboration workspaces.

\subsection{AI agent} \label{sec:methods:ai}

\paragraph{Context.} \label{sec:methods:ai:context}
The platform uses OpenAI's multimodal \texttt{gpt-4o} model (specifically \texttt{gpt-4o-2024-08-06}), which processes text and images in a single model. To give the AI full context of the task and collaboration, each API call to \texttt{gpt-4o} is prompted with the information on the screen so it has the full context of the collaboration. In the prompt, we include the same text of the task given to participants, previous submissions, current copy, elapsed time, history of its own actions, history of its chain-of-thought, chat history, and general instructions. Moreover, a screenshot is taken of the image after every change and included as an input so the AI can observe the image and its evolution over time and edits. These details enable human collaboration with a fully functional, multimodal AI agent. The prompt template is shown in Appendix~\ref{app:prompt_agent}.

\paragraph{Actions.} \label{sec:methods:ai:actions}
To ensure that the human-human and human-AI conditions are comparable, the AI agent can take the same actions as a human participant except for submitting ads. The actions include sending messages, editing any element of the ad copy (including the headline, primary text, and description), selecting images, generating images using the Dall-E 3 interface, and waiting (\textit{i.e.}, not taking any action). The agent is prompted every 10 seconds whether to engage in action.

\paragraph{Chain-of-thought.} \label{sec:methods:ai:cot}
To ensure the agent is taking actions appropriately, we used chain-of-thought prompting \citep{wei2022chainofthought}.\footnote{We used OpenAI's structured output feature for chain-of-thought prompting.} Specifically, we prompt the model with questions to reflect on the state of the collaboration. These questions were necessary to avoid undesirable model behavior, such as repeatedly sending the same message and were developed and perfected through trial and error.

\subsection{Procedure} \label{sec:methods:procedure}

\begin{figure}[!ht]
    \includegraphics[width=0.95\linewidth]{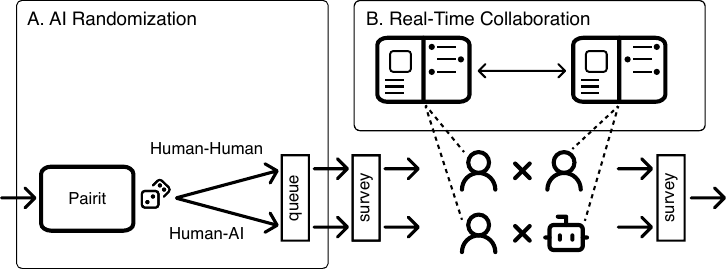}
    \caption{\small Overview of methods. (A) Participants are randomized into collaborating with another participant or an AI agent. (B) Participants collaborate with another participant or an AI agent to produce ads in a real-time collaborative workspace.}
    \label{fig:overview}
\end{figure}

\paragraph{AI randomization and queuing.} As soon as a participant was redirected to our platform from Prolific, they were randomized into either the human-human or the human-AI condition (Figure~\ref{fig:overview}A). In the human-human condition, participants joined a queue until another participant was available, at which point they were paired with each other. In the human-AI condition, a participant joined a simulated queue, in which they waited for a random amount of time between 1 and 5 seconds, after which they were paired with an AI agent. We do not reveal whether or not the partner is a human or an AI until the post-task survey.

\paragraph{Pre-task survey.} 
After participants were paired with each other, they answered a 10-item survey, on a 7-point Likert scale, to measure their Big Five personality traits \citep{rammstedt2007personality}. 

\paragraph{Ad creation task.} Once the participants were paired and completed a pre-task survey, they entered the collaborative workspace in which they could message each other and edit and submit ads (Figure~\ref{fig:overview}B). The edits were synchronized and messages were transmitted between the participants in real-time using websockets (\textit{i.e.}, \href{https://tiptap.dev/}{tiptap.dev} and \href{https://pusher.com/}{pusher.com}), as they are in commercial collaboration tools (\textit{e.g.}, Google Docs) and chat applications (\textit{e.g.}, Slack). The participants had 40 minutes to submit as many ads as they could produce. At the end of the 40 minutes, the participants were automatically redirected to the post-task survey.

\paragraph{Post-task survey.} After the completion of the ad creation task, the participants answered questions regarding their perception of AI, all on a 7-point Likert scale. These included two questions about their experience using AI (\textit{i.e.}, ``I have used artificial intelligence (AI) chatbots before (e.g., ChatGPT, Bard).'' and ``I had a positive experience using AI chatbots.''), one question about their perception of their partner as an AI (\textit{i.e.}, ``I believe my partner was an AI during the task.'') and a final question in which we revealed the identity of their partner (human or AI) and asked whether that changed the perception of the quality of the collaboration (\textit{i.e.}, ``Your partner was [an AI assistant/a human]. Knowing this, to what extent has your perception of the quality of your collaboration changed?''). 85\% of participants completed the post-task survey.

\subsection{Participants} \label{sec:methods:participants}

We pre-registered a sample of 2,500 participants from Prolific (\href{https://prolific.com/}{www.prolific.com}) based in the US, with representative stratification across gender and ethnicity. In total, 2,310 participants completed the task. Of the 2,500 participants who entered, 23 did not enter the matching queue (see Section~\ref{sec:methods:procedure} for details) and were removed from the study. A further 167 participants quit the study or were timed out before they were matched to a partner. In 83 human-human sessions (14\% of HH teams), the matched partner did not complete the task due to dropout or timeout. Because these sessions do not represent collaborative work and submissions required confirmation by both participants, we exclude them from all analyses, yielding a final sample of 2,234 participants (1,258 human-AI, 976 human-human). The overall attrition rate was 7.6\%. The study was conducted from October 15 to 18, 2024, and took a median of 46.32 minutes to complete. The participants were paid \$9.

\subsection{Summary statistics and randomization checks} \label{sec:methods:randomization}

In total, the dataset includes 2,234 participants, 1,751 teams, 11,024 display ad submissions, 182,607 messages, 62,119 edits on images, 1,889,559 text edits, and 10,074 AI-generated images. To ensure that the randomization procedure successfully balanced covariates across experimental conditions, Table~\ref{tb:randomization_check} compares key participant characteristics between the human-human and human-AI conditions. These include demographic variables (e.g. age, gender) and participants' work status, and Appendix~\ref{app:bigfive} adds the Big Five personality traits. No significant differences were detected across treatment conditions, except a slight difference in neuroticism that does not survive a correction for multiple tests, indicating balanced randomization across participants.

\begin{table}[ht]
\centering
\begin{threeparttable}
\begin{tabular}{lccccc}
\hline \\[-4ex]
\textit{Variables} & \textit{All} & \textit{Human-AI} & \textit{Human-Human} & \textit{t-statistic (SE)} & \textit{p-value} \\
\hline \\[-4ex]
Individuals & 2234 & 1258 & 976 & -- & -- \\
Teams & 1751 & 1258 & 493 & -- & -- \\
Gender (\% Male) & 50.8\% & 50.4\% & 51.3\% & -0.438 (0.021) & 0.661 \\
Age & 42 $\pm$ 14 & 43 $\pm$ 15 & 42 $\pm$ 14 & 1.296 (0.608) & 0.195 \\
Full-Time & 47.7\% & 48.5\% & 46.7\% & 0.830 (0.021) & 0.407 \\
Part-Time & 14.6\% & 13.4\% & 16.2\% & -1.883 (0.015) & 0.060 \\
\hline \hline
\end{tabular}
\begin{tablenotes}
\small
\item Notes: $^{*}$p$<$0.05, $^{**}$p$<$0.01, $^{***}$p$<$0.001. The level of analysis is the individual participant. Observations represent the $N=2,234$ randomized participants. For t-statistics, gender is coded as 1 for male and 0 for female, and employment status is 1 for the listed category. Balance checks for Big Five personality traits appear in Table~\ref{tb:A:bigfive}.
\end{tablenotes}
\caption{Randomization checks.} \label{tb:randomization_check}
\end{threeparttable}
\end{table}

\subsection{Incentives} \label{sec:methods:incentives}

To incentivize participants to create high-quality ads, we informed them that they were eligible for additional \$100 prizes based on the quantity and quality of the ads they submitted, as well as the performance of the ads in the marketing field experiment. This incentive structure was identical across human-human and human-AI conditions. Participants were explicitly instructed that ``the greater the number of ads, the greater your chances, but not if the ads are of low quality.'' Ultimately, two participants were awarded \$100 each.

\subsection{Message labeling} \label{sec:methods:message_label}

To label messages into categories, we prompt \texttt{gpt-4o-mini-2024-07-18} for each message independently and ask for a category label. We enforce that the labels are from the set of pre-determined labels using OpenAI's Structured Outputs API to ensure schema compliance. The classification scheme consists of six categories: (1) \textit{content} (messages sharing information, facts, or deliverables directly related to the task), (2) \textit{process} (messages addressing strategies, approaches, or real-time logistical details for the session), (3) \textit{social} (messages building rapport or containing social interactions unrelated to the task), (4) \textit{emotional} (messages expressing emotions or attitudes related to the session or task), (5) \textit{feedback} (messages providing constructive feedback or evaluative comments), and (6) \textit{other} (messages that do not fit into the other categories). We group \textit{content} and \textit{process} messages as \textit{task-oriented} communication, and \textit{social} and \textit{emotional} messages as \textit{interpersonal} communication, while \textit{feedback} and \textit{other} remain separate. See Appendix~\ref{app:prompt_message_label} for the exact prompt and Python code.

\subsection{Evaluation of ad quality} \label{sec:methods:eval}

\paragraph{Participants.}
To obtain human ratings of ad quality, we recruited a separate sample of 1,168 raters from Prolific based in the US, with representative stratification across gender, age, and ethnicity. We built a custom platform for this survey, run on Google Cloud Platform's App Engine.\footnote{The code is available on \href{https://github.com/harangju/pairit-survey-mockup/}{GitHub}.} This evaluation was conducted from November 7 to 9, 2024, and took a median of 16.47 minutes to complete.

\paragraph{Ad samples.} 
To obtain ratings for all ads while avoiding survey fatigue, we created a random sample, with random order, of 40 ads per participant. To target at least 3 ratings per ad, we produced a set of 1,300 samples. As a participant entered our survey platform, we drew one sample from the set, without replacement, to provide to the participant, with each participant receiving a unique sample.

\paragraph{Ad mockups.}
To obtain ratings of display ads that are as close as possible to display ads one would see on digital ad publishers, we first created mockups of each display ad. An ad mockup is a simulated representation of a display ad designed to closely resemble the appearance and functionality of an actual ad as it would appear on digital ad publishers' platforms. We built a web application that populates the image, ad copy, shortened link, call-to-action, and other user interface items, including the \textit{Like}, \textit{Comment}, \textit{Share}, and \textit{Close} buttons, the profile picture, and the \textit{Sponsored} tag. Screenshots were then programmatically taken of each mockup.

\paragraph{Survey items.}
We used mockups of display ads. For each ad, we asked participants to provide three ratings regarding the quality of the text, the image, and the estimated click through rate. Each item was rated on a 7-point Likert scale. The first rating was based on the following prompt ``The text is present, clear, relevant, and engaging''; the second on ``The image is visually appealing''; and the third on ``I am likely to click on this ad.'' See Figure~\ref{fig:survey} for an example of an item on the survey.

\subsection{Field evaluation of ad performance}

To assess real-world advertising outcomes, we ran ad campaigns on the social media platform X, pre-registered on \href{https://osf.io/95dhu}{osf.io/95dhu}. We use the following as outcome variables: click-through rate (CTR), cost-per-click (CPC), view-through rate (VTR; as a fraction of document viewed), and view-through duration (VTD; in seconds). We tracked VTR and VTD by using a unique link on DocSend for each ad.

Due to the effects of divergent targeting in A-B testing \citep{braun2025abtesting}, we follow recommendation 5 from \cite{braun2024adplatforms} and run a multi-ad study to test the causal effects of the ads on performance, conditional on online algorithms. We do not use a holdout in this study. Each ad has a unique DocSend link, so we expect the outcome in a hypothetical holdout group to be zero and attribute all document visits to the ads \citep{braun2024adplatforms}.

To run as many ads as possible while sampling evenly from human-human and human-AI teams and from the human predictions of ad quality, we implemented a stratified sampling approach. We obtained a sample of 2,000 ads, sampling between one and two ads from each participant. We first divided the ads into those created by human-human teams and human-AI teams and then further divided the ads into 10 strata according to the human predictions of the likelihood of user clicks, from low to high predicted click probability (see Section~\ref{sec:methods:eval}). We had already removed 8 ads that potentially violated moderation policies (due to the inclusion of \textit{e.g.} violence, sexual, or drug content). We controlled for advertising spend in our analyses to account for auto-bidding. Consistent with other analyses, field analyses exclude 37 ads from human-human sessions where one partner did not complete the collaboration, and one ad for which X reported no data.

To prevent overlapping audience targeting, we assigned five unique ads to each of 400 campaigns, structuring them as 5-ad split tests within individual campaigns (noting the platform's limit of five split tests per campaign). To further ensure no overlap across campaigns, we allocated a random set of 24 to 25 ZIP codes to each campaign, selecting ZIP codes with populations between 10,000 and 100,000.\footnote{The data is from \href{https://data.census.gov/table/ACSDT5Y2020.B01003}{2020 American Community Survey}.} We tested the robustness of this allocation across ZIP codes by conducting one-way ANOVA on population and income, yielding non-significant results: $F(399, 9354) = 0.954$, $p = 0.734$ for population, and $F(399, 9354) = 0.973$, $p = 0.636$ for income.

Ad impressions were delivered from January 21, 2025, to February 9, 2025. Due to platform limits in running separate ad campaigns, we ran 50 ad campaigns for two days at a time. We control for any potential temporal confounders with campaign random effects. 

\subsection{Model specifications} \label{sec:methods:model}

\paragraph{Human-AI collaboration effects.}
Due to the experimental nature of our design, we measure the effects of human-AI collaboration using a standard regression model:
\begin{equation} \label{eq:hai}
    Y_i = \delta \text{HAI}_i + \beta X_i + \epsilon_i \text{,}
\end{equation}
where $Y_i$ is the outcome, HAI$_i$ is 1 if the unit is in the human-AI condition and 0 otherwise, $X_i$ includes demographic controls (age and gender), and $\epsilon_i$ is the error term. The unit of analysis $i$ varies by outcome: individual-level for messages, copy edits, image edits, AI image generations, submissions, message categories, and delegation rates; team-level for team submissions; and rating-level for ad quality scores. Team-level analyses include no controls.

\paragraph{Field evaluations.}
To evaluate the results of our field experiment, we estimate three mixed-effects specifications for each field outcome. Specification (1) includes only the Human-AI indicator and spend. Specification (2) includes only the human-rated text and image quality measures and spend. Specification (3) is the joint model:
\begin{equation} \label{eq:field}
\begin{aligned}
    Y_{tc} = & \ \delta \text{HAI}_{tc}
    + \beta_1 \text{Text}_{tc} + \beta_2 \text{Image}_{tc} \\
    &+ \gamma \text{Spend}_{tc} + u_c + \epsilon_{tc} \text{,}
\end{aligned}
\end{equation}
where $Y_{tc}$ represents the outcome (CPC, CTR, or VTD) for ad $t$ in campaign $c$, $\text{Image}$ and $\text{Text}$ are human-rated quality measures from Section~\ref{sec:methods:eval}, $\text{Spend}_{tc}$ is the ad's spend, and $u_c \sim N(0, \sigma^2_u)$ is the random intercept for campaign $c$. For CTR, we fit a binomial mixed model of clicks out of impressions with the same predictors. We exclude the human-rated click-likelihood measure because it is an outcome proxy that stands in for realized CTR. Empirically, click likelihood correlates highly with text and image ratings ($r = 0.55$ and $0.68$, versus $0.34$ between text and image). Alternative models that include click likelihood do not alter our conclusions (Table~\ref{tb:A:field_click}).

\paragraph{Process analysis.}
To relate collaboration processes to outcomes, we estimate regressions using Equation~\ref{eq:hai}, substituting the relevant predictor and outcome. We regress productivity, quality, and output diversity on communication patterns (\textit{i.e.}, task-oriented, interpersonal) and on delegation (Table~\ref{tb:mechanism}). This approach identifies how communication patterns and delegation relate to outcomes.

\paragraph{Recognition of partner identity.}
To test whether correct identification of partner type moderates treatment effects, we estimate:
\begin{equation} \label{eq:recognition}
\begin{aligned}
    Y_i = &\beta_0 + \beta_1 \text{HAI}_i + \beta_2 \text{Recognition}_i \\
    &\quad + \beta_3 (\text{HAI}_i \times \text{Recognition}_i) + \gamma X_i + \varepsilon_i \text{.}
\end{aligned}
\end{equation} A significant $\beta_3$ indicates that the treatment effect differs depending on whether participants correctly identified their partner type.

To construct a measure of recognition, we use the partner perception question (``I believe my partner was an AI during the task''), measured on a 7-point Likert scale from strongly disagree to strongly agree. Using the scale midpoint (score $\geq$ 4 = ``believed AI''), we define \textit{Recognition} = 1 if participants correctly identified their partner type: believed AI in the H-AI treatment, or believed human in the H-H treatment. Recognition analyses include only participants who completed the post-task survey.

\paragraph{Delegation.} \label{sec:methods:delegation}
We measure delegation as the fraction of text editing work performed by the partner rather than the focal participant. We defined delegation as one minus the participant's character contribution divided by the session's total character contribution, which, on average for human-human teams, is mechanically 50\%. For human-AI teams, we detect AI-generated content by identifying large text jumps (simultaneous edits of more than 10 characters), which is how the AI produced the content compared to the human editing incrementally.

\paragraph{Output diversity.} \label{sec:methods:diversity}
To measure output diversity, we embed each submission's ad copy (headline, primary text, and description) using OpenAI's \texttt{text-embedding-3-small} model. We compute the centroid (mean embedding vector) for each treatment group and calculate the cosine distance from each submission to its own treatment's centroid. Higher values indicate greater distance from the typical output in that condition, reflecting more diverse or atypical content. We aggregate to the user level by averaging across submissions and estimate treatment effects using Equation~\ref{eq:hai}. This analysis is restricted to users who submitted at least one ad with text content ($N$=2,034 of 2,234 participants).

\section{Results}\label{sec:results}

\subsection{Productivity, Quality and Output Diversity}\label{sec:results:outcomes}

We begin by examining the primary outcomes of human-AI collaboration---productivity, ad quality, and output diversity---and the collaboration processes that enable them. Human-AI teams send 56\% more messages, make 73\% fewer direct edits, and delegate 17\% more work (Table~\ref{tb:collaboration}). These processes accompany higher productivity per-worker, higher ad text quality, lower ad image quality and more homogeneous (less diverse) outputs (Table~2).

\paragraph{Collaborating with AI agents increases individual-level productivity.}
Participants were incentivized to submit as many ads as possible within the time limit. Participants in human-AI teams submitted 50\% more ads per worker, per unit time than their counterparts in human-human teams, with results consistent when demographic controls are added (Table~\ref{tb:A:productivity}). AI collaboration thus supports individual productivity by enabling participants to generate more output per worker compared to those in human-human teams.

\begin{table}[ht]
\centering
\begin{threeparttable}
\begin{tabular}{lccccc}
\hline
 & \textit{Productivity} & \multicolumn{3}{c}{\textit{Quality}} & \textit{Output} \\[-3pt]
\cmidrule(lr){3-5}\addlinespace[-7pt]
 & & \textit{Text} & \textit{Image} & \textit{Click} & \textit{Diversity} \\
\hline
\textbf{Intercept} & 4.601$^{***}$ & 4.686$^{***}$ & 4.801$^{***}$ & 3.709$^{***}$ & 0.303$^{***}$ \\
 & (0.346) & (0.042) & (0.037) & (0.037) & (0.010) \\
\textbf{Human-AI} & 1.954$^{***}$ & 0.323$^{***}$ & -0.136$^{***}$ & -0.014 & -0.094$^{***}$ \\
 & (0.197) & (0.024) & (0.021) & (0.021) & (0.006) \\
\hline
Demographics & Yes & Yes & Yes & Yes & Yes \\
Observations & 2234 & 45578 & 45578 & 45578 & 2034 \\
\hline \hline
\end{tabular}
\begin{tablenotes}
\small
\item Notes: $^{*}$p$<$0.05, $^{**}$p$<$0.01, $^{***}$p$<$0.001. The level of analysis varies by outcome: participant level for \textit{Productivity} ($N=2,234$ participants) and \textit{Output Diversity} ($N=2,034$ participants with text submissions); rating level for \textit{Text}, \textit{Image}, and \textit{Click} ($N=45,578$ ratings clustered at the ad level). \textit{Productivity} is the number of submissions by participants. \textit{Text}, \textit{Image}, \textit{Click} are human ratings (7-point scale) with standard errors clustered at the ad level. \textit{Output Diversity} is cosine distance to treatment centroid (higher is more diverse). Robust standard errors account for heteroskedasticity. Full ratings model with covariates appears in Table~\ref{tb:A:human_ratings_full}.
\end{tablenotes}
\caption{Collaborating with AI agents improves productivity and text quality but lowers image quality (\textit{i.e.}, a \textit{jagged frontier}), and decreases output diversity.} \label{tb:outcomes}
\end{threeparttable}
\end{table}

It is important to note that our study did not include a human-alone condition. Prior studies have consistently shown that AI tools enhance individual productivity compared to humans working alone across various tasks \citep{noy2023experimental, brynjolfsson2025genai, dellacqua2026navigating, chen2024large, wiles2025writing}. In light of this evidence, our results suggest that the AI in human-AI teams acts as a partial substitute for an additional human collaborator. Without a human-alone condition, however, we cannot directly quantify AI's incremental contribution beyond what an individual human might have achieved working alone. We explore this limitation further in the Discussion section.

With that caveat in mind, to test the robustness of our results to the alternative hypothesis that individual-level productivity differences are simply due to the submission of incomplete ads, we examined ad copy completion rates across conditions. As shown in Table~\ref{tb:A:completion}, participants in human-AI teams had consistently higher completion rates for ad copy elements compared to participants in human-human teams. Taken together, these findings suggest that human-AI teams are indeed more productive, in both the number of ads submitted and the ad completion rate.

\paragraph{Collaborating with AI agents improves text quality but lowers image quality.} In human evaluations of the quality of the ads, we found that human-AI teams produced higher quality ad text but lower quality ad images compared to human-human teams (Table~\ref{tb:outcomes}). The estimated likelihood of clicking on the ad was indistinguishable between the two groups. The human ratings reveal that AI introduces important trade-offs in output quality across different media formats. Specifically, collaborating with LLM-based agents improves text quality but reduces image quality. In our setting, LLM-based partners excel at next-word text generation but struggle to judge which visual elements are eye-catching or appealing. In human-human teams, peer discussions and disagreements allow partners to negotiate visual choices and produce images that both find acceptable. Human-AI teams may struggle to replicate this collaborative evaluation process. Because AI is strong at text but not image evaluation, human participants might place less trust in the AI's image suggestions. This potential trust asymmetry could help explain why human-AI image quality falls even as text quality rises. The contrast echoes the \textit{jagged frontier} pattern in which AI gains vary across tasks \citep{dellacqua2026navigating}. These results were supported by AI evaluations of ad quality (see Table~\ref{tb:A:ai_ratings_full}).\footnote{Interestingly, in our AI evaluations of ad quality, we found that AI ratings were higher on the text and clicks for ads produced by human-AI teams and the same across the groups for image quality. In a way, it is unsurprising that the AI rated these ads' text quality, image quality, and estimated click likelihood as equal to or better than those produced by human-human teams because these ads were created in collaboration with OpenAI's \texttt{gpt-4o}.} These results suggest that while LLMs enhance text-based outputs, their contributions to multimodal outputs like ad images may require complementary tools or fine-tuning for image-related tasks. 

\paragraph{Collaborating with AI agents reduces output diversity.} Beyond productivity and quality, we examined whether collaboration with AI agents affects output diversity. To measure output diversity, we compute the cosine distance of each submission's text embedding to its respective treatment centroid, which represents the mean embedding vector of all ad copy within that experimental condition (see Section~\ref{sec:methods:diversity} for details on the embedding model). A higher cosine distance indicates that a submission is more atypical or diverse relative to other ads in the same condition. As shown in Table~\ref{tb:outcomes}, human-AI teams produce outputs that are closer to their condition centroid than human-human teams ($\beta=-0.094$, $p<0.001$), indicating a homogenization of team outputs \citep{chen2024large, padmakumar2024doeswritinglanguagemodels}. This diversity collapse suggests that while AI-generated content may be higher quality on average on certain dimensions, it lacks the variation that human creativity provides.

\subsection{Ad Performance in the Field}\label{sec:results:effectiveness}

Our findings on productivity and quality raise an important question about whether and how these trade-offs meaningfully impact ad performance. To evaluate real-world performance of ads created by human-human and human-AI teams, we conducted a field experiment on X, a social media platform, generating 4,932,373 impressions and 7,546 clicks over 20 days. We examined how human-AI collaboration and ad quality influenced key advertising metrics, including cost-per-click (CPC), click-through rate (CTR), and view-through duration (VTD) (Table~\ref{tb:field}).

\begin{table}[ht]
\centering
\begin{threeparttable}
\setlength{\tabcolsep}{4pt}
\begin{tabular}{@{}c@{}}\resizebox{\linewidth}{!}{\begin{tabular}{lccccccccc}
\hline \\[-4ex]
 & \multicolumn{3}{c}{\textit{CPC (\$)}} & \multicolumn{3}{c}{\textit{CTR (log-odds)}} & \multicolumn{3}{c}{\textit{VTD (log-sec)}} \\
\cline{2-4} \cline{5-7} \cline{8-10}
 & (1) & (2) & (3) & (1) & (2) & (3) & (1) & (2) & (3) \\
\hline \\[-4ex]
\textbf{Intercept} & 10.237$^{***}$ & 11.564$^{***}$ & 11.495$^{***}$ & -6.600$^{***}$ & -6.808$^{***}$ & -6.799$^{***}$ & 0.412$^{***}$ & 0.100 & 0.108 \\
 & (0.330) & (0.657) & (0.679) & (0.022) & (0.074) & (0.076) & (0.069) & (0.122) & (0.124) \\
\textbf{Human-AI} & 0.116 &  & 0.111 & -0.011 &  & -0.021 & -0.000 &  & -0.014 \\
 & (0.213) &  & (0.214) & (0.027) &  & (0.027) & (0.046) &  & (0.047) \\
\textbf{Image} &  & -0.180$^{\dagger}$ & -0.175 &  & 0.008 & 0.008 &  & -0.001 & -0.001 \\
 &  & (0.109) & (0.110) &  & (0.012) & (0.012) &  & (0.021) & (0.021) \\
\textbf{Text} &  & -0.069 & -0.076 &  & 0.034$^{*}$ & 0.035$^{*}$ &  & 0.054$^{**}$ & 0.055$^{**}$ \\
 &  & (0.098) & (0.098) &  & (0.014) & (0.014) &  & (0.020) & (0.021) \\
\hline \\[-4ex]
Spend & Yes & Yes & Yes & Yes & Yes & Yes & Yes & Yes & Yes \\
Campaign RE & Yes & Yes & Yes & Yes & Yes & Yes & Yes & Yes & Yes \\
Observations & 1822 & 1822 & 1822 & 1962 & 1962 & 1962 & 1675 & 1675 & 1675 \\
\hline \hline
\end{tabular}}\end{tabular}
\begin{tablenotes}
\small
\item Notes: $^{\dagger}$p$<$0.1, $^{*}$p$<$0.05, $^{**}$p$<$0.01, $^{***}$p$<$0.001. The level of analysis is the ad for \textit{CPC} ($N=1,822$ ads with at least one click), \textit{CTR} ($N=1,962$ ads), and \textit{VTD} ($N=1,675$ ads with at least one DocSend view), where VTD averages log view time over each ad's views. For each dependent variable, column (1) estimates the treatment effect of the Human-AI collaboration indicator; column (2) estimates the effects of human-rated text and image quality; and column (3) estimates the joint model. Spend is controlled in all specifications to account for auto-bidding, and we include campaign random effects. Standard errors are clustered by campaign. CPC excludes ads with zero clicks.
\end{tablenotes}
\caption{Field experiment results show no baseline differences between collaboration types, but reveal associations between specific quality characteristics and ad performance.} \label{tb:field}
\end{threeparttable}
\end{table}

\paragraph{Click measures}

We evaluate click performance (CPC and CTR) with campaign random effects models. For CPC, column (1) shows that the Human-AI collaboration treatment has no statistically significant relationship with CPC (0.116, or about 1.5\% of mean CPC of \$7.53, $p > 0.05$).\footnote{Higher spend consistently lowers costs, consistent with divergent targeting under auto-bidding. We therefore interpret the effects in this section as conditional on ad algorithms, following recommendation 5 from \cite{braun2024adplatforms}.} In column (2), we estimate the effects of ad quality measures on performance. Higher image quality is marginally associated with lower CPC ($-0.180$ dollars, or about 2.4\% of mean CPC, $p < 0.1$). Under the joint model in column (3), the coefficient on the Human-AI indicator remains small and statistically insignificant (0.111, $p > 0.05$), while the coefficient on image quality is similar in size but no longer marginally significant ($-0.175$, $p > 0.1$). Text quality does not significantly predict CPC in either specification. For CTR, column (1) reveals no significant baseline performance difference between treatment groups ($-0.011$ log-odds, $p > 0.05$). In column (2), each one-point increase in text quality is associated with about 3.5\% higher CTR (0.034 log-odds, $p < 0.05$), relative to a mean CTR of 0.14\%. Under the joint model in column (3), both coefficients remain virtually identical. Specifically, the Human-AI treatment effect remains null, and the positive relationship between text quality and CTR remains robust (0.035, $p < 0.05$). Image quality does not significantly relate to CTR.

\paragraph{View measures}

To measure consumer attention in the field, we analyze view-through duration (VTD; in log-seconds) and view-through rate (VTR), each averaged over the DocSend views of an ad. In column (2), each one-point increase in text quality is associated with about 5.5\% longer view time (0.054 log-seconds, $p < 0.01$). Under the joint model in column (3), that coefficient remains similar (0.055, $p < 0.01$), while the Human-AI treatment effect is negligible ($-0.014$, $p > 0.05$). Column (1) likewise shows no baseline VTD difference between human-AI and human-human ads. Image quality does not predict VTD in either model. When we instead estimate VTD per view, which weights each ad by its number of views, the text coefficient is smaller and insignificant (0.024, $p > 0.05$; Table~\ref{tb:A:views}). In auxiliary analyses, we find no significant treatment or quality effects on VTR (Table~\ref{tb:A:views}).

\paragraph{Human-AI collaboration raises CTR and view time through text quality.}
We test whether text and image quality mediate the effect of human-AI collaboration on each field outcome to evaluate how uneven quality gains influence ad performance despite a null baseline difference. The \textit{jagged frontier} implies that human-AI collaboration shifts quality unevenly, raising text quality while reducing image quality (Table~\ref{tb:outcomes}). When downstream outcomes respond unevenly to those components, a null total effect can conceal a real gain, and indirect effects can operate even when the total effect is null \citep{zhao2010reconsidering}. We estimate both quality pathways in a multiple-mediator model to isolate each dimension's indirect effect \citep{preacher2008asymptotic}. Because the joint model includes campaign random effects, we compute 95\% Monte Carlo confidence intervals from 200,000 joint draws of both coefficients rather than refitting the models in bootstrap samples \citep{preacher2012advantages}. Human-AI ads have higher text quality (0.322, $p<0.001$) and lower image quality ($-0.114$, $p<0.05$) on the 7-point scale (Table~\ref{tb:A:field_mediation}). Through text quality, human-AI collaboration raises CTR by about 1\% (0.011 log-odds, 95\% CI [0.002, 0.022]) and lengthens view time by about 2\% (0.020 log-seconds, 95\% CI [0.005, 0.038]). The pathways through image quality are close to zero for CTR and VTD. On CPC, the two pathways run in opposite directions: lower image quality raises CPC (0.020) while higher text quality lowers it ($-0.026$), so the two quality effects offset, although neither is statistically significant on its own. 

\paragraph{The Jagged Frontier of AI Capabilities Obscures Human-AI Collaboration Performance.} Across all specifications, ads from human-AI and human-human teams perform similarly in the field. Text quality predicts CTR and view time, and image quality is marginally associated with CPC, but the baseline collaboration-type effect is null. The direct effects of human-AI collaboration are small and insignificant ($-0.021$ for CTR and $-0.014$ for VTD, both $p>0.05$), a pattern that \citet{zhao2010reconsidering} call indirect-only mediation. While human-AI collaboration improves ad quality on dimensions where AI capabilities are strong (i.e. text), it marginally reduces ad quality on dimensions where AI capabilities are relatively weaker (i.e. images), and the net result is that ads produced by human-human and human-AI teams perform similarly, on average. Taken together, these results imply average human-AI teams do not outperform average human-human teams because jagged AI capabilities both improve and reduce performance along different dimensions. 

\subsection{Communication and Delegation Patterns}\label{sec:results:processes}

Having established the outcomes of human-AI collaboration, we now examine differences in the collaboration process itself. Teams collaborate through two primary modes: communication and task execution. We analyze both to understand the shifts in teamwork dynamics underlying the productivity, quality and performance outcomes.

\paragraph{Collaborating with AI agents increases overall communication.}
To test how human-AI collaboration affects communication, we measured the number of messages sent by participants in both human-human and human-AI groups. Participants worked collaboratively with and sent messages to their partners in real-time, for both human-human and human-AI groups. In the human-human group, participants messaged their human partners; in the human-AI group, participants messaged the AI. The Pairit platform logs a timestamped record of each message sent by a participant or AI. Participants in human-AI teams sent 56\% more messages than those in human-human teams, with consistent results when demographic covariates were included (Table~\ref{tb:collaboration}). These results indicate that collaborating with an AI partner encourages more frequent communication overall.

\begin{table}[ht]
\centering
\begin{threeparttable}
\begin{tabular}{lccccc}
\hline
 & \multicolumn{3}{c}{\textit{Communication}} & \multicolumn{2}{c}{\textit{Task Execution}} \\[-3pt]
\cmidrule(lr){2-4} \cmidrule(lr){5-6}\addlinespace[-7pt]
 & \textit{Count} & \textit{Task-Oriented} & \textit{Interpersonal} & \textit{Copy Edits} & \textit{Delegation} \\
\hline
\textbf{Intercept} & 20.296$^{***}$ & 0.362$^{***}$ & 0.580$^{***}$ & 1676.458$^{***}$ & 0.478$^{***}$ \\
 & (1.450) & (0.014) & (0.015) & (57.646) & (0.018) \\
\textbf{Human-AI} & 12.580$^{***}$ & 0.090$^{***}$ & -0.107$^{***}$ & -1032.406$^{***}$ & 0.083$^{***}$ \\
 & (0.843) & (0.008) & (0.008) & (33.588) & (0.010) \\
\hline
Demographics & Yes & Yes & Yes & Yes & Yes \\
Observations & 2234 & 2234 & 2234 & 2234 & 2233 \\
\hline \hline
\end{tabular}
\begin{tablenotes}
\small
\item Notes: $^{*}$p$<$0.05, $^{**}$p$<$0.01, $^{***}$p$<$0.001. The level of analysis is the individual participant. Observations represent the $N=2,234$ participants (with $N=2,233$ for \textit{Delegation}). \textit{Count} is the number of messages sent by a participant. \textit{Task-oriented} is the fraction of messages labeled as \textit{content} or \textit{process}. \textit{Interpersonal} is the fraction of messages labeled as \textit{emotional} or \textit{social}. \textit{Copy edits} are counts of text changes (\textit{i.e.}, text input events). \textit{Delegation} is fraction of text work performed by partner. Robust standard errors account for heteroskedasticity.
\end{tablenotes}
\caption{Collaborating with AI agents increases communication, with a higher share of task-oriented and a lower share of interpersonal messages. It also decreases direct edits to the ad copy while increasing delegation to AI agents.} \label{tb:collaboration}
\end{threeparttable}
\end{table}

\paragraph{Collaborating with AI agents shifts teams from \textit{interpersonal} toward \textit{task-oriented} communication.}
In addition to the number of messages sent across conditions, we investigated whether the content of messages varied across conditions. We used \texttt{gpt-4o-mini} to label each message into categories, which we grouped into \textit{task-oriented} and \textit{interpersonal} communication (see Section~\ref{sec:methods:message_label}). We found that human-AI teams sent a 24\% higher share of task-oriented messages, while human-human teams sent a 23\% higher share of interpersonal messages (Table~\ref{tb:collaboration}). This shift indicates that collaboration with AI emphasizes task-oriented over interpersonal interaction, possibly because participants can focus more on the task without needing to navigate the social or emotional aspects typical of human collaboration.

\paragraph{Collaborating with AI agents shifts work from direct editing to delegation.}
Participants in human-AI teams made 73\% fewer direct copy edits compared to human-human teams (Table~\ref{tb:collaboration}). However, participants in the human-AI condition still engaged in direct editing, indicating that while AI collaboration reduced the frequency of manual edits, it did not eliminate them entirely. Human-AI teams also delegated 58\% of work to AI partners compared to 50\% in human-human teams ($\beta=0.083$, $p<0.001$), a 17\% increase. Unlike previous studies, such as \cite{chen2024large}, which artificially constrain participants to specific modalities when interacting with AI, our results show that in more unconstrained collaborative settings, where humans and agents can freely allocate effort to different tasks, participants utilized both direct involvement with the work product (ad editing) and interaction through messaging.

Taken together, these patterns in communication and task execution demonstrate how AI partners elicit more directive, less interpersonal communication and associated task handoff, resulting in a \textit{delegation workflow}. Within this workflow, humans direct and delegate tasks to their AI partners, communicating more frequently and with greater task focus while reducing direct manual editing.

\paragraph{Session interaction dynamics.}
We track granular interaction logs within sessions to analyze how team communication and editing behavior evolve over time (Appendix~\ref{app:interaction_dynamics}, Figure~\ref{fig:session_trajectories}, and Tables~A9--A10). Human-AI teams diverge early from human-human teams in message rates, edit rates, and task-oriented communication, and the gap persists through the session. From the outset, human-AI teams send more messages and make fewer direct edits. Within sessions, positive verbal appraisals of the AI's work increase local delegation. In session fixed-effects models with $\pm$3-minute windows, the partner's share of text characters increases by 6.6 percentage points and the partner's writing rate by 27.4 partner characters per minute ($p<0.001$ for both; Table~A11). The same appraisals do not significantly increase partner share or writing rate in human-human teams. Human-AI teams thus appear to discover their partner's capabilities and then delegate, navigating the jagged frontier.

\paragraph{Illustrative examples.}
To ground these quantitative findings, we extracted representative message exchanges. In H-AI teams, communication is directive: ``Can you generate an image with a professional looking person at a desk?'' followed by the AI responding with a generated image. Coordination appears as explicit task allocation: ``Let's divide this up. I'll work on the headline and you handle the image.'' In contrast, H-H teams show rapport-building: ``Sorry for the delay! Had to grab coffee'' ... ``No worries, take your time.'' H-H teams also engage in collaborative negotiation: ``What do you think about using blue instead?'' ... ``Hmm, I was thinking green might work better because...'' ... ``Good point, let's try both and see.'' These exchanges illustrate the directive versus collaborative dynamics that distinguish human-AI from human-human teams.

\subsection{Teamwork and Collaboration Dynamics}\label{sec:results:mechanisms}

We analyze team communication and task delegation as collaboration processes, and how partner recognition moderates collaboration behavior (Section~\ref{sec:methods:model}). We find that participants who correctly recognize their AI partner engage in more task-oriented and less interpersonal communication with their AI partner. We also find that both communication style and delegation are associated with output quality.

First, participants in the human-AI treatment group were highly accurate in identifying their partner as AI (94.5\%), while human-human group participants were relatively less accurate (57.4\%). When we tested whether this recognition shaped collaboration behavior (Equation~\ref{eq:recognition}) we found that correct identification predicts communication volume and style (Table~\ref{tb:recognition}). Participants who correctly identified their partner as an AI agent sent more messages ($\beta_3=11.510$, $p<0.001$), with a higher proportion of task-oriented messages ($\beta_3=0.081$, $p<0.05$) and a lower proportion of interpersonal messages ($\beta_3=-0.093$, $p<0.01$). They also delegated somewhat more ($\beta_3=0.083$, $p<0.10$). 

We find that recognition of the AI partner acts as an amplifier of adaptation. Because nearly all human-AI participants recognized their partner, we cannot determine whether it is a necessary condition (Appendix Table A13). Correct recognition is associated with stronger adaptations (Table~\ref{tb:recognition}). Recognition thus moderates and enhances the shift toward directive communication. Explicit identity probes during chat are uncommon but associated with higher post-session recognition (Table~A14).

\begin{table}[ht]
\centering
\begin{threeparttable}
\begin{tabular}{lccccc}
\hline
 & \multicolumn{3}{c}{\textit{Communication}} & \multicolumn{2}{c}{\textit{Task Execution}} \\[-3pt]
\cmidrule(lr){2-4} \cmidrule(lr){5-6}\addlinespace[-7pt]
 & \textit{Count} & \textit{Task-Oriented} & \textit{Interpersonal} & \textit{Copy Edits} & \textit{Delegation} \\
\hline
\textbf{Intercept} & 17.564$^{***}$ & 0.369$^{***}$ & 0.585$^{***}$ & 1624.753$^{***}$ & 0.492$^{***}$ \\
 & (1.657) & (0.017) & (0.018) & (75.444) & (0.021) \\
\textbf{Human-AI $\times$} & 11.510$^{***}$ & 0.081$^{*}$ & -0.093$^{**}$ & -40.898 & 0.083$^{\dagger}$ \\
\quad\raisebox{4pt}{\textbf{Recognition}} & (2.728) & (0.034) & (0.035) & (85.406) & (0.047) \\
\textbf{Human-AI} & 2.404 & 0.029 & -0.039 & -1029.573$^{***}$ & 0.023 \\
 & (2.507) & (0.033) & (0.033) & (69.461) & (0.046) \\
\textbf{Recognition} & 4.724$^{***}$ & -0.017 & 0.013 & 119.888$^{\dagger}$ & -0.019$^{\dagger}$ \\
 & (1.044) & (0.013) & (0.013) & (64.889) & (0.011) \\
\hline
Demographics & Yes & Yes & Yes & Yes & Yes \\
Observations & 1906 & 1902 & 1902 & 1906 & 1905 \\
\hline \hline
\end{tabular}
\begin{tablenotes}
\small
\item Notes: $^{\dagger}$p$<$0.1, $^{*}$p$<$0.05, $^{**}$p$<$0.01, $^{***}$p$<$0.001. The level of analysis is the individual participant. Observations represent the $N=1,906$ participants who answered the recognition question (with $N=1,902$ for message fractions and $N=1,905$ for \textit{Delegation}). \textit{Recognition} indicates whether the participant correctly identified their partner type. \textit{Count} is the number of messages sent by a participant. \textit{Task-oriented} is the fraction of messages labeled as \textit{content} or \textit{process}. \textit{Interpersonal} is the fraction of messages labeled as \textit{emotional} or \textit{social}. \textit{Copy edits} are counts of text changes (\textit{i.e.}, text input events). \textit{Delegation} is fraction of text work performed by partner. Robust standard errors account for heteroskedasticity.
\end{tablenotes}
\caption{Recognition of the AI partner moderates communication.} \label{tb:recognition}
\end{threeparttable}
\end{table}

Second, communication style and delegation predict productivity, ad quality and output diversity, but these effects vary across human-AI and human-human teams (Table~\ref{tb:mechanism}, Panel A). Base coefficients report human-human associations; Human-AI interaction terms report how those associations change in human-AI teams. In human-human teams, task-oriented communication is negatively associated with productivity ($\beta=-4.19$, $p<0.001$), text quality ($\beta=-0.420$, $p<0.05$), and image quality ($\beta=-0.278$, $p<0.05$). The Human-AI interactions reverse these patterns for task-oriented communication ($\beta=4.69$, $p<0.001$ for productivity; $\beta=0.750$, $p<0.001$ for text; $\beta=0.516$, $p<0.01$ for image; $\beta=0.585$, $p<0.001$ for click), implying positive within-human-AI associations with quality. In contrast, interpersonal communication is positively associated with productivity in human-human teams ($\beta=4.25$, $p<0.001$), while the Human-AI interactions are negative ($\beta=-4.80$, $p<0.001$ for productivity; $\beta=-0.689$, $p<0.001$ for text; $\beta=-0.589$, $p<0.001$ for image; $\beta=-0.601$, $p<0.001$ for click), implying negative within-human-AI associations with quality. Our analysis suggests that the performance payoff of communication style depends on partner type. Interpersonal communication is associated with higher productivity in human-human teams, perhaps through rapport-building and coordination, whereas task-oriented, directive communication predicts higher performance in human-AI teams.

\begin{table}[ht]
\centering
\begin{threeparttable}
\begin{tabular}{lccccc}
\hline
 & \textit{Productivity} & \multicolumn{3}{c}{\textit{Quality}} & \textit{Output} \\[-3pt]
\cmidrule(lr){3-5}\addlinespace[-7pt]
 & & \textit{Text} & \textit{Image} & \textit{Click} & \textit{Diversity} \\
\hline
\rowcolor{gray!10} \multicolumn{6}{l}{\textit{Panel A: Communication}} \\
\hline
\textbf{Task-Oriented} & -4.187$^{***}$ & -0.420$^{*}$ & -0.278$^{*}$ & -0.288$^{*}$ & 0.011 \\
 & (0.847) & (0.165) & (0.128) & (0.124) & (0.027) \\
\textbf{Task-Oriented $\times$ H-AI} & 4.693$^{***}$ & 0.750$^{***}$ & 0.516$^{**}$ & 0.585$^{***}$ & -0.035 \\
 & (1.242) & (0.206) & (0.174) & (0.166) & (0.035) \\
\textbf{Interpersonal} & 4.247$^{***}$ & 0.273$^{\dagger}$ & 0.219$^{\dagger}$ & 0.196$^{\dagger}$ & 0.014 \\
 & (0.754) & (0.156) & (0.119) & (0.118) & (0.026) \\
\textbf{Interpersonal $\times$ H-AI} & -4.799$^{***}$ & -0.689$^{***}$ & -0.589$^{***}$ & -0.601$^{***}$ & 0.030 \\
 & (1.141) & (0.195) & (0.168) & (0.161) & (0.034) \\
\hline
Demographics & Yes & Yes & Yes & Yes & Yes \\
Observations & 2041 & 2041 & 2041 & 2041 & 2030 \\
\hline
\rowcolor{gray!10} \multicolumn{6}{l}{\textit{Panel B: Task Execution}} \\
\hline
\textbf{Delegation} & -1.382$^{*}$ & 0.302$^{***}$ & 0.132$^{\dagger}$ & 0.178$^{**}$ & -0.083$^{***}$ \\
 & (0.586) & (0.071) & (0.069) & (0.065) & (0.012) \\
\hline
Demographics & Yes & Yes & Yes & Yes & Yes \\
Observations & 1222 & 1222 & 1222 & 1222 & 1222 \\
\hline \hline
\end{tabular}
\begin{tablenotes}
\small
\item Notes: $^{\dagger}$p$<$0.1, $^{*}$p$<$0.05, $^{**}$p$<$0.01, $^{***}$p$<$0.001. The level of analysis is the individual participant. Observations represent the $N=2,041$ participants in Panel A (with $N=2,030$ for output diversity) who sent at least one message and had at least one rated ad, and $N=1,222$ human-AI participants in Panel B. Panel A: both treatments; base coefficients are H-H. Panel B: H-AI only. \textit{Task-oriented} is the fraction of messages labeled as \textit{content} or \textit{process}. \textit{Interpersonal} is the fraction of messages labeled as \textit{emotional} or \textit{social}. \textit{Delegation} is fraction of text work performed by partner. \textit{Productivity} is the number of submissions by participants. \textit{Text}, \textit{Image}, \textit{Click} are human ratings (7-point scale). \textit{Output Diversity} is cosine distance to treatment centroid (higher is more diverse). Robust standard errors account for heteroskedasticity.
\end{tablenotes}
\caption{Communication style and task execution predict performance differently in human-AI vs. human-human teams.} \label{tb:mechanism}
\end{threeparttable}
\end{table}

In human-AI teams, delegation is associated with higher text ratings ($\beta=0.302$, $p<0.001$) and click ratings ($\beta=0.178$, $p<0.01$) but does not significantly predict image ratings ($\beta=0.132$, $p<0.1$; Table~\ref{tb:mechanism}, Panel B; see also Table~A12), consistent with the \textit{jagged frontier} of AI capabilities \citep{dellacqua2026navigating}. Click ratings likely track text quality because clicking is partially driven by copy appeal, and delegation is associated with higher text quality without worsening image quality. We do not interpret this asymmetric pattern as evidence of wrong or suboptimal delegation decisions by human participants. Our measure captures text handoff, so the pattern shows that text handoff predicts text and click ratings but not image ratings, consistent with task-capability matching along the jagged frontier. Delegation also predicts lower output diversity ($\beta=-0.083$, $p<0.001$), tracking the diversity collapse observed in human-AI teams. Notably, delegation is associated with lower productivity ($\beta=-1.38$, $p<0.05$), perhaps because heavy delegators are less engaged overall. In summary, recognition of the AI partner is associated with a more task-oriented communication style and somewhat more delegation, which track variations in ad quality and diversity without proving a causal decomposition of structure versus capability.

Third, human-AI teams that employ work and communication processes which complement AI, perform better, on average, and transcend the limitations of the jagged frontier of AI capabilities. For example, human-AI teams that engage in task-oriented communication produce higher productivity, higher quality ads across all dimensions and display no diversity collapse in their collaborative outputs. This implies that human-AI teams with AI appropriate communication processes transcend the jagged frontier and avoid output homogenization while generating higher ad quality and productivity across the board, even on metrics, like image quality, on which AI is inferior on average. In contrast, human-AI teams that engaged in more interpersonal communication were less productive and produced lower quality ads across all dimensions, failing to take advantage of capabilities, like text generation, which are within the jagged frontier and on which AI is most capable.

\section{Discussion}\label{sec:Discussion}

\subsection{Theoretical, managerial, and methodological implications}

\paragraph{Theoretical implications.}

This study applies established teamwork process concepts to human-AI collaboration in an open-dialogue, agentic creative setting, demonstrating how LLM-based AI agents alter work processes, communication dynamics, and task execution in collaborative settings \citep{schneider2021agent}. Our findings connect to three effects documented in the AI literature: productivity gains \citep{noy2023experimental, brynjolfsson2025genai}, the \textit{jagged frontier} of AI capabilities \citep{dellacqua2026navigating}, and diversity collapse \citep{padmakumar2024doeswritinglanguagemodels, chen2024large}. We find that human-AI collaboration induces a \textit{delegation workflow}---task-oriented, directive communication and increased delegation to AI---which is associated with higher text quality (an AI strength) but not image quality (an AI weakness), with treatment effects shaped by the jagged frontier. Delegation also reduces output diversity, as teams that delegate more produce outputs more similar to the typical human-AI output. This extends theoretical work on human-AI collaboration design \citep{agarwal2025sufficient} and behavioral trust in AI teammates \citep{zhang2023teammate} from decision-making contexts to agentic, multimodal creative work. 

The shift from interpersonal to task-oriented communication aligns with work showing that humans apply social behaviors to machines \citep{reeves1996media}, but that these behaviors differ systematically from human-human interactions in that people interacting with AI become more demanding and instrumental \citep{tey2024judge}. This shift reflects adaptation to AI partners that follow instructions reliably and require no social maintenance. While AI can streamline team processes \citep{wilson2018collaborative}, our findings revealed trade-offs. In our experiment, use of AI enhanced ad text quality but reduced image quality.

The field results also have implications for how researchers and organizations evaluate AI. Ads from human-AI and human-human teams performed similarly overall, yet human-AI collaboration increased click-through rates and view time through higher text quality, and weaker images did not undo this advantage. Total effects alone would suggest that AI made no difference to campaign performance. As AI capabilities are jagged, however, AI alters distinct components of work performance (sometimes in opposite directions) rather than lifting performance across the board. When downstream outcomes depend on those components unevenly, a null net effect can conceal a real gain along a single dimension. In our setting, AI's text advantage increased click-through rates and view time even though overall performance did not differ, showing that the \textit{jagged frontier} extends beyond rated quality into consumer behavior. Evaluations of AI deployments that rely only on aggregate outcomes may miss gains that run through specific capabilities. Theories of human-AI collaboration must therefore account not only for task-specific competencies, but also for how those competencies translate to downstream outcomes \citep{dellacqua2026navigating}.

Critically, we find that \textit{recognition}, or participants' awareness that their partner is an AI, acts as an amplifier of these collaboration dynamics. Participants who correctly identified their AI partner exhibited stronger shifts toward task-oriented communication, suggesting that explicit awareness of AI capabilities amplifies behavioral adaptation. Because nearly all participants recognized their AI partner, our data cannot determine whether these patterns would arise without recognition. The asymmetry in recognition accuracy (94.5\% in human-AI teams versus 57.4\% in human-human teams) suggests that AI communication and task behaviors are highly distinctive and detectable, triggering explicit trust calibration and behavioral adaptation.

Moreover, our work challenges the traditional conceptualization of AI as a passive tool or assistant \citep{noy2023experimental, dellacqua2026navigating, chen2024large} by positioning LLM-based agents as active collaborators that shape team dynamics and, in turn, productivity and output quality \citep{anthony2023, makarius2020rising}. For example, the AI's proficiency in generating ad copy influenced participants to focus on instructions rather than direct edits. Rather than merely augmenting human efforts, AI agents co-create outcomes, introducing new dynamics that affect team performance.

Our findings also speak to the literature on team cognition and distributed expertise \citep{collins2024building}. In human-human teams, expertise is often shared through interpersonal exchanges, which foster trust but can slow decision-making. In contrast, human-AI teams in our study bypassed much of this social overhead. In our setting, coordination becomes more directive than mutual, conflict is avoided rather than resolved, and task allocation looks more like delegation than negotiation. These patterns describe how humans adapt communication and handoff when collaborating with AI in this creative-production task.

\paragraph{Managerial implications.}
Our work also offers actionable insights for managers and organizations seeking to integrate AI agents into collaborative workflows, particularly in creative and productivity-driven environments. Our findings demonstrate that AI agents significantly reduce social coordination costs, allowing team members to focus on task execution and content generation. Managers in industries like marketing, content creation, or consulting can leverage AI agents to streamline routine tasks, such as drafting campaign materials or reports, freeing human employees for strategic activities like audience analysis or brand positioning.

However, the study also highlights critical limitations that managers must address to fully harness AI's potential in multimodal workflows. While LLM-based agents excelled in text generation, they underperformed in image-related tasks. The text advantage still paid off in the market, raising click-through rates and view time even though overall campaign performance did not differ, so managers who judge AI only by aggregate metrics may overlook gains from tasks within AI capabilities. This trade-off supports the idea that organizations should adopt a hybrid approach along the \textit{jagged frontier} of AI capabilities \citep{dellacqua2026navigating}, allocating tasks within AI capabilities to agents and those outside AI capabilities, such as image selection, to humans or fine-tuning models to perform better on those tasks. Such an approach can mitigate the quality disparities observed in our study and is particularly relevant in domains where compelling visual elements drive downstream outcomes.

Furthermore, our results offer guidance on structuring human-AI teams to optimize productivity and performance. AI's ability to handle bulk editing tasks allowed participants to engage in more content ideation and quality control. Managers can apply this division of labor to design efficient workflows, assigning AI agents to repetitive or data-intensive tasks while reserving human expertise for creative oversight and final approvals. To implement such workflows, organizations should invest in training programs that teach employees how to delegate tasks to AI effectively, fostering a collaborative culture that maximizes both technological and human contributions.

Finally, the study underscores the importance of aligning AI deployment with organizational goals and team dynamics. Our findings reveal that higher delegation to AI reduces output diversity, a concern for organizations that value creative variation. To mitigate diversity collapse, managers could design workflows that preserve human-driven ideation, such as human-only brainstorming sessions before AI-assisted execution. Although human-AI teams produce fewer total submissions than human-human teams, the 50\% increase in individual productivity suggests that AI can serve as a partial substitute for additional personnel in certain contexts. Managers could use this insight to scale team output without proportionally increasing headcount, a cost-effective strategy for startups or resource-constrained firms. However, they must also be mindful of ``collapsing onto the jagged frontier'': as teams rely more on AI, they may improve at tasks where AI excels (text generation) while atrophying at tasks where AI underperforms (image selection). Organizations should ensure that human skills in domains in which AI is weak are maintained through deliberate practice or task rotation.

\paragraph{Methodological implications.}

This study advances the methodological landscape for studying human-AI collaboration through the development of the Pairit platform, a novel experimental platform that redefines how researchers investigate team dynamics in real-time, task-driven settings. Unlike traditional approaches that rely on aggregate performance metrics or isolated AI interactions \citep{noy2023experimental, dellacqua2026navigating}, Pairit captures granular, time-stamped data on every keystroke, message, edit, and API call, enabling a detailed reconstruction of collaborative workflows. In our experiment, this capability allowed us to analyze 182,607 messages, 1,889,559 text edits, 62,119 image edits, and 10,074 AI-generated images, revealing nuanced patterns such as how AI-driven suggestions shifted human effort from direct text editing to instructional messaging. This methodological advance allowed us to analyze both aggregate and granular dynamics of human-AI teamwork and offers researchers new opportunities to discover insights regarding human-AI decision-making, task allocation, and interaction processes.

The randomization of team composition represents another significant methodological advance, moving beyond human-only versus human-AI comparisons to include human-human and human-AI pairings. Unlike prior research that often contrasts AI-augmented individuals with those working alone \citep{noy2023experimental, vaccaro2024combinations}, our approach allowed us to evaluate the effects on collaboration itself, which is critical to understand given the proliferation of AI agents in workplaces. This randomization design is also adaptable to other configurations, such as teams with multiple humans or multiple AI agents, enabling researchers to test how team composition influences outcomes in contexts like project management, educational group work, or distributed software development.

To fairly compare human-human and human-AI collaboration, another critical contribution is that AI agents are \textit{agentic}. That is, AI agents on Pairit are capable of performing nearly all actions available to humans, including sending chat messages, editing text, selecting images, and generating new images via external APIs (e.g., DALL-E 3), with the sole exception of final ad submission in this study. This approach ensures direct comparability between human-AI and human-human conditions, a departure from prior studies where AI is often just a chatbot \citep{chen2024large, brynjolfsson2025genai}. By enabling AI to act independently and actively, we analyze how AI's active participation alters collaborative dynamics, which would not have been possible in experimental designs that treat AI as a passive tool.

Pairit's multimodal capabilities further distinguish it as a methodological tool, supporting tasks that integrate text, images, and external API interactions. While our study focused on ad creation, Pairit's architecture is extensible to diverse domains, such as software development (e.g., AI generating code while humans debug), scientific writing (e.g., AI drafting literature reviews), or multimedia content production (e.g., designing virtual reality environments). This versatility allows researchers to explore human-AI collaboration in tasks requiring varied cognitive and creative skills, addressing a gap in prior studies that often limit AI to text-based functions \citep{brynjolfsson2025genai}.

These methodological contributions have broader implications for experimental research and organizational practice. Pairit's ability to simulate realistic collaborative environments offers a testing ground for refining theoretical models of human-AI interaction and informing the design of AI-driven workflows. Ultimately, Pairit's methodological innovations bridge the gap between controlled experiments and real-world applications and provide a robust, scalable platform for advancing our understanding of human-AI collaboration.

\subsection{Limitations and future directions} \label{sec:discussion:limitations}

\paragraph{Limitations.}
While we conducted one of the first truly agent-randomized experiments in human-AI collaboration, combining laboratory and field tests, on a newly developed platform capable of recording fine-grained collaboration dynamics, our work is not without limitations. First, our design documents human-AI versus human-human process patterns and treats recognition as a measured moderator, leaving the direct experimental manipulation of identity labels and independent variation of AI capability to future research. We cannot fully disentangle whether these patterns stem from the nature of AI communication or from human participants adapting their behavior upon recognizing their partner as AI. Our analyses tracing the pathway from recognition to communication to delegation to quality are \textit{post hoc} and correlational; while they are consistent with a causal pathway, alternative explanations remain possible. For instance, participants who are more comfortable with AI may both recognize it more readily and adopt more efficient collaboration strategies, with no causal relationship between recognition and behavior. Second, the limitations of current AI models, particularly vision models, present challenges for multimodal tasks. These models are optimized for next-word generation and specific visual tasks, such as identifying items in images, but are not designed for nuanced assessments like image quality prediction. This limitation likely contributed to the lower image quality observed in human-AI teams and underscores the need for AI systems tailored to specific creative and evaluative tasks. However, AI is a moving target and image selection and design competency likely will improve over time. As with any study of AI, results may drift as AI changes. Moreover, while our study provides a controlled experimental context, it may not fully capture the complexities of long-term collaboration with AI agents in real-world environments. Future research should explore how these dynamics evolve over extended periods and in more diverse task domains to validate and expand on our findings.

Finally, our study lacks a human alone condition, which prevents us from directly assessing how human-AI teams compare to individuals working independently. Without this baseline, we cannot precisely isolate AI's marginal contribution to productivity and the coordination costs associated with collaboration. Nevertheless, our findings have practical implications for organizations looking to integrate AI agents into collaborative workflows, particularly in tasks requiring creativity and coordination. Our work lays a foundation for understanding human-AI collaboration in complex tasks, and future research can build on this by incorporating a human-alone condition to further disentangle marginal productivity effects.

Regarding generalizability, our findings should apply to tasks that share key features with ad creation: multi-modal creative tasks requiring both text and visual elements, where AI capabilities vary across modalities (the \textit{jagged frontier}); iterative production tasks involving multiple rounds of generation, evaluation, and refinement; and settings with objective quality metrics. Our findings may be less applicable to high-stakes decision tasks where errors are costly and irreversible or tasks requiring deep domain expertise where AI capabilities are more limited. While our study captured within-session adaptation (humans learning AI capabilities through multiple ad iterations), longitudinal dynamics across repeated collaborations remain unexplored. The core process patterns we document (task-oriented communication, higher delegation, and capability-dependent quality effects) should generalize broadly, but the \textit{direction} of quality effects will depend on the specific task and AI capabilities.

\paragraph{Future directions.}
Our work opens several promising avenues for future research. First, ours is one of the first randomized controlled trials to evaluate how AI agents affect work and communication processes. In our detailed, task-level analysis, we uncovered specific effects on communication, collaboration, and work processes, but these findings represent only a first step. Future work should provide more task-level evidence about how collaboration with AI agents changes communication, teamwork, work processes, and organization design, and how these effects shape productivity and performance across task and organizational contexts. Such research is essential to understanding the impact of AI on work and organizational performance. Within that agenda, two designs would sharpen identification beyond what partner-type randomization with measured belief can deliver. A labeled versus unlabeled design would measure the effect of telling people their partner is AI. A design that varies AI capability while holding partner type fixed would separate collaboration patterns from task capability. Second, the limitations of current vision models suggest a need for separate, specialized image generation models. Future studies could incorporate state-of-the-art visual models designed specifically for tasks like image quality prediction and generation, enabling a more nuanced understanding of AI contributions to multimodal creative tasks. Third, our findings are grounded in the context of ad design, but they invite exploration across a range of other collaborative domains. Important contexts include software development (e.g., coding), data analysis, collaborative writing, and financial accounting. Investigating human-AI collaboration in these areas could reveal domain-specific dynamics and inform the broader design of AI systems tailored to various professional workflows. Finally, extending this research to longitudinal settings could provide insights into how human-AI collaboration evolves over time. Long-term studies could examine the sustainability of productivity gains, the development of trust, and the potential for ``learning effects'' where humans adapt to working with AI agents or vice versa. These investigations would bridge the gap between controlled experimental settings and real-world applications, enhancing the generalizability of our findings. We intend to make Pairit freely available to academics advancing peer-reviewed research into human-AI collaboration.

\subsection{Conclusion}

Our work provides novel insights into how AI agents alter work processes, productivity, and performance in an open-dialogue, agentic setting where humans could have explored further and partners could act on shared work. Our findings connect to three effects documented in the AI literature. First, human-AI teams achieved 50\% higher productivity per worker, suggesting that AI agents can partially substitute for human collaborators. Second, we observed the \textit{jagged frontier} of AI capabilities: text quality improved in human-AI teams while image quality declined, and in the field, higher text quality increased the click-through rates and view time of human-AI ads even though overall performance did not differ. Third, we found diversity collapse: teams that delegated more to AI produced outputs more similar to the typical human-AI output. 

We then found two collaboration processes associated with these effects: \textit{task orientation} (human-AI teams sent a 24\% higher share of task-oriented messages and a 19\% lower share of interpersonal messages) and \textit{delegation} (participants delegated 17\% more work to AI and made 73\% fewer direct edits). AI-appropriate work processes, in turn, improved performance in human-AI teams and transcended the jagged frontier of AI capabilities. For example, task-oriented communication produced higher productivity, higher quality ads across all dimensions, and displayed no diversity collapse in their outputs, while human-AI teams with more interpersonal communication were less productive and produced lower quality ads across all dimensions. Participants who correctly recognized their partner as an AI were even more task-oriented and less interpersonal. In contrast to human-AI teams, in human-human teams, interpersonal communication improved productivity. Delegation improved text quality in human-AI teams and was positively associated with diversity collapse, creating homogeneous outputs of higher average quality. 

These findings suggest that AI agents drive productivity, performance and output diversity by reshaping teamwork along the jagged frontier of AI capabilities and imply that future work should investigate the work and communication processes that complement AI and unlock the productivity and performance gains in human-AI collaboration.

%
%
%

\section*{Acknowledgments}

We thank Dean Eckles and John Horton for their invaluable discussions. This work was supported by the Initiative on the Digital Economy at the Sloan School of Management at the Massachusetts Institute of Technology. The study was approved by the Massachusetts Institute of Technology institutional review board. The code and data are available upon request.

\bibliography{references}

\clearpage
\appendix

\setcounter{table}{0}
\renewcommand{\thetable}{A\arabic{table}}
\setcounter{figure}{0}
\renewcommand{\thefigure}{A\arabic{figure}}
\renewcommand{\theHtable}{A\arabic{table}}
\renewcommand{\theHfigure}{A\arabic{figure}}
\renewcommand{\theHsection}{A\arabic{section}}

\section{Big Five Personality Balance Checks}\label{app:bigfive}

To complement the demographic balance checks reported in Table~\ref{tb:randomization_check}, we examined whether Big Five personality traits were balanced across experimental conditions. Table~\ref{tb:A:bigfive} shows no significant differences in openness, conscientiousness, extraversion, or agreeableness between human-human and human-AI conditions. Neuroticism differs slightly ($p = 0.022$) but not after a Bonferroni correction for five tests.

\begin{table}[ht]
\centering
\begin{threeparttable}
\begin{tabular}{lccccc}
\hline \\[-4ex]
\textit{Trait} & \textit{All} & \textit{Human-AI} & \textit{Human-Human} & \textit{t-statistic (SE)} & \textit{p-value} \\
\hline \\[-4ex]
Openness & 0.72 $\pm$ 0.19 & 0.71 $\pm$ 0.19 & 0.72 $\pm$ 0.19 & -0.642 (0.008) & 0.521 \\
Conscientiousness & 0.79 $\pm$ 0.16 & 0.79 $\pm$ 0.16 & 0.79 $\pm$ 0.16 & -0.794 (0.007) & 0.427 \\
Extraversion & 0.56 $\pm$ 0.21 & 0.55 $\pm$ 0.20 & 0.56 $\pm$ 0.21 & -0.821 (0.009) & 0.412 \\
Agreeableness & 0.71 $\pm$ 0.19 & 0.70 $\pm$ 0.19 & 0.71 $\pm$ 0.18 & -0.122 (0.008) & 0.903 \\
Neuroticism & 0.49 $\pm$ 0.22 & 0.50 $\pm$ 0.22 & 0.48 $\pm$ 0.22 & 2.299 (0.009) & 0.022 \\
\hline \hline
\end{tabular}
\begin{tablenotes}
\small
\item Notes: $^{*}$p$<$0.05, $^{**}$p$<$0.01, $^{***}$p$<$0.001. Personality traits are normalized from a 7-point Likert scale.
\end{tablenotes}
\caption{Big Five personality trait balance checks.} \label{tb:A:bigfive}
\end{threeparttable}
\end{table}

\section{Productivity}\label{app:productivity}

Table~\ref{tb:A:productivity} reports the full regression results for productivity outcomes at both team and individual levels. At the team level, human-AI teams produced fewer total submissions, which is expected given that these teams have half as many human workers. At the individual level, however, participants in human-AI teams submitted 50\% more ads than their counterparts in human-human teams.

\begin{table}[ht]
\centering
\begin{threeparttable}
\begin{tabular}{lccc}
\hline \\[-4ex]
 & \textit{Team} & \multicolumn{2}{c}{\textit{Individual}} \\
\hline \\[-4ex]
\textbf{Intercept} & 7.623$^{***}$ & 3.850$^{***}$ & 4.601$^{***}$ \\
 & (0.277) & (0.129) & (0.346) \\
\textbf{Human-AI} & -1.847$^{***}$ & 1.925$^{***}$ & 1.954$^{***}$ \\
 & (0.315) & (0.197) & (0.197) \\
\hline \\[-4ex]
Demographics & No & No & Yes \\
Observations & 1751 & 2234 & 2234 \\
\hline \hline
\end{tabular}
\begin{tablenotes}
\small
\item Notes: $^{*}$p$<$0.05, $^{**}$p$<$0.01, $^{***}$p$<$0.001. Robust standard errors account for heteroskedasticity.
\end{tablenotes}
\caption{Productivity of human-AI and human-human teams.} \label{tb:A:productivity}
\end{threeparttable}
\end{table}

\section{Work Actions}\label{app:work_actions}

Table~\ref{tb:A:work_actions} presents detailed results on work actions across experimental conditions. Participants in human-AI teams made substantially fewer direct copy edits but more image edits, suggesting a shift from direct text manipulation to delegation and visual content creation.

\begin{table}[ht]
\centering
\begin{threeparttable}
\begin{tabular}{lccc}
\hline \\[-4ex]
 & \textit{Copy Edits} & \textit{Image Edits} & \textit{AI Images} \\
\hline \\[-4ex]
\textbf{Intercept} & 1676.458$^{***}$ & 28.397$^{***}$ & 6.067$^{***}$ \\
 & (57.646) & (1.559) & (0.390) \\
\textbf{Human-AI} & -1032.406$^{***}$ & 6.592$^{***}$ & 0.328 \\
 & (33.588) & (0.931) & (0.219) \\
\hline \\[-4ex]
Demographics & Yes & Yes & Yes \\
Observations & 2234 & 2234 & 2234 \\
\hline \hline
\end{tabular}
\begin{tablenotes}
\small
\item Notes: $^{*}$p$<$0.05, $^{**}$p$<$0.01, $^{***}$p$<$0.001. Robust standard errors account for heteroskedasticity.
\end{tablenotes}
\caption{Collaborating with AI reduces copy edits but increases image edits at the individual level.} \label{tb:A:work_actions}
\end{threeparttable}
\end{table}

\section{Message Categories}\label{app:message_categories}

Table~\ref{tb:A:message_categories} reports the distribution of message categories across experimental conditions. Messages were classified into content, process, social, emotional, feedback, and other categories using \texttt{gpt-4o-mini}. Human-AI teams sent significantly more content- and process-oriented messages, while human-human teams sent more social and emotional messages.

\begin{table}[ht]
\centering
\begin{threeparttable}
\begin{tabular}{lccccc}
\hline \\[-4ex]
 & \textit{Content} & \textit{Process} & \textit{Emotional} & \textit{Social} & \textit{Feedback} \\
\hline \\[-4ex]
\textbf{Intercept} & 0.177$^{***}$ & 0.185$^{***}$ & 0.192$^{***}$ & 0.387$^{***}$ & 0.047$^{***}$ \\
 & (0.012) & (0.010) & (0.010) & (0.013) & (0.005) \\
\textbf{Human-AI} & 0.050$^{***}$ & 0.041$^{***}$ & -0.043$^{***}$ & -0.064$^{***}$ & 0.023$^{***}$ \\
 & (0.006) & (0.006) & (0.005) & (0.007) & (0.003) \\
\hline \\[-4ex]
Demographics & Yes & Yes & Yes & Yes & Yes \\
Observations & 2234 & 2234 & 2234 & 2234 & 2234 \\
\hline \hline
\end{tabular}
\begin{tablenotes}
\small
\item Notes: $^{*}$p$<$0.05, $^{**}$p$<$0.01, $^{***}$p$<$0.001. Robust standard errors account for heteroskedasticity.
\end{tablenotes}
\caption{Collaborating with AI increases content-, process-, and feedback-related messages while decreasing social and emotional messages.} \label{tb:A:message_categories}
\end{threeparttable}
\end{table}

\section{Ad Copy Completion}\label{app:completion}

Table~\ref{tb:A:completion} examines whether productivity differences might be driven by submission of incomplete ads. Participants in human-AI teams had consistently higher completion rates for all ad copy elements (headline, primary text, and description), indicating that the productivity gains reflect genuine output rather than incomplete submissions.

\begin{table}[ht]
\centering
\begin{threeparttable}
\begin{tabular}{lccc}
\hline \\[-4ex]
 & \textit{Headline} & \textit{Primary Text} & \textit{Description} \\
\hline \\[-4ex]
\textbf{Intercept} & 0.808$^{***}$ & 0.769$^{***}$ & 0.719$^{***}$ \\
 & (0.025) & (0.026) & (0.027) \\
\textbf{Human-AI} & 0.150$^{***}$ & 0.169$^{***}$ & 0.161$^{***}$ \\
 & (0.014) & (0.014) & (0.015) \\
\hline \\[-4ex]
Demographics & Yes & Yes & Yes \\
Observations & 2234 & 2234 & 2234 \\
\hline \hline
\end{tabular}
\begin{tablenotes}
\small
\item Notes: $^{*}$p$<$0.05, $^{**}$p$<$0.01, $^{***}$p$<$0.001. Robust standard errors account for heteroskedasticity.
\end{tablenotes}
\caption{Individuals in human-AI teams submit ads with more copy completed.} \label{tb:A:completion}
\end{threeparttable}
\end{table}

\section{Survey Interface}\label{app:survey}

\begin{figure}[ht]
    \centering
    \includegraphics[width=0.7\linewidth]{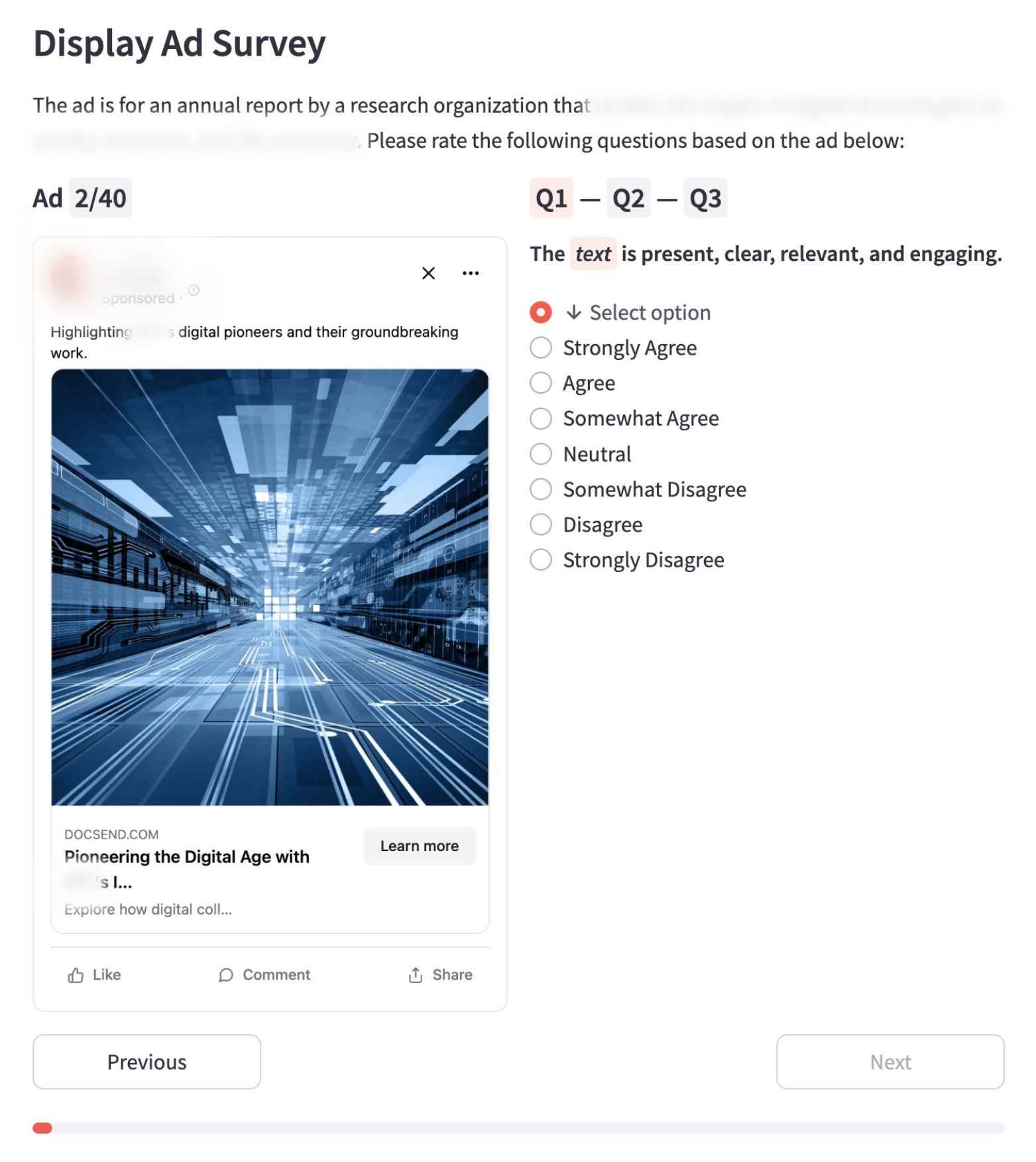}
    \caption{The user interface for the ad quality survey.}
    \label{fig:survey}
\end{figure}

Figure~\ref{fig:survey} shows the user interface for the ad quality survey described in Section~\ref{sec:methods:eval}.

\clearpage
\section{Human Ratings with Full Covariates}\label{app:human_ratings_full}

Table~\ref{tb:A:human_ratings_full} presents the complete regression results for human quality ratings, including demographic controls and work process covariates.

\begin{table}[ht]
\centering
\begin{threeparttable}
\begin{tabular}{lcccccc}
& \multicolumn{2}{c}{\textit{Text}} & \multicolumn{2}{c}{\textit{Image}} & \multicolumn{2}{c}{\textit{Click}} \\
\hline \\[-4ex]
\textbf{Intercept} & 4.686$^{***}$ & 4.628$^{***}$ & 4.801$^{***}$ & 4.581$^{***}$ & 3.709$^{***}$ & 3.507$^{***}$ \\
& (0.042) & (0.061) & (0.037) & (0.051) & (0.037) & (0.053) \\
\textbf{Human-AI} & 0.323$^{***}$ & 0.417$^{***}$ & -0.136$^{***}$ & -0.167$^{***}$ & -0.014 & 0.028 \\
& (0.024) & (0.053) & (0.021) & (0.044) & (0.021) & (0.045) \\
\hline \\[-4ex]
\textbf{\# Submissions} &  & -0.029$^{***}$ &  & -0.012$^{***}$ &  & -0.018$^{***}$ \\
    &  & (0.002) &  & (0.001) &  & (0.001) \\
\textbf{\# AI Images Generated} &  & 0.008$^{***}$ &  & 0.029$^{***}$ &  & 0.023$^{***}$ \\
    &  & (0.002) &  & (0.002) &  & (0.002) \\
\textbf{\# Human Messages} &  & 0.004$^{***}$ &  & 0.002$^{***}$ &  & 0.002$^{***}$ \\
    &  & (0.000) &  & (0.000) &  & (0.000) \\
\textbf{\# Human Copy Edits} &  & 0.137$^{***}$ &  & 0.041$^{**}$ &  & 0.081$^{***}$ \\
    &  & (0.017) &  & (0.015) &  & (0.015) \\
\textbf{\# Human Copy Edits $\times$ H-AI} &  & -0.028 &  & -0.001 &  & -0.020 \\
    &  & (0.027) &  & (0.027) &  & (0.026) \\
\textbf{\# Human Image Edits} &  & -2.328$^{**}$ &  & -1.065 &  & -0.644 \\
    &  & (0.897) &  & (0.710) &  & (0.728) \\
\textbf{\# Human Image Edits $\times$ H-AI} &  & 2.448$^{*}$ &  & 2.089$^{*}$ &  & 1.813$^{*}$ \\
    &  & (1.009) &  & (0.829) &  & (0.834) \\
\hline \\[-4ex]
Demographics & Yes & Yes & Yes & Yes & Yes & Yes \\
Observations & 45578 & 45578 & 45578 & 45578 & 45578 & 45578 \\
\hline \hline \vspace{0.1em}
\end{tabular}
\begin{tablenotes}
\small
\item Notes: $^{*}$p$<$0.05, $^{**}$p$<$0.01, $^{***}$p$<$0.001. Standard errors are clustered at the ad level. Copy and image edits are per thousand.
\end{tablenotes}
\caption{Human ratings of ads with full covariates.} \label{tb:A:human_ratings_full}
\end{threeparttable}
\end{table}

\clearpage
\section{AI Ratings with Full Covariates}\label{app:ai_ratings_full}

\paragraph{AI ratings.}
To obtain AI ratings of ad quality, we prompt OpenAI's multimodal \texttt{gpt-4o-mini-2024-07-18}, which supports image input and structured outputs.\footnote{See OpenAI's documentation on \href{https://platform.openai.com/docs/guides/vision}{vision} and \href{https://platform.openai.com/docs/guides/structured-outputs/}{structured outputs}.} To make AI evaluations comparable to human evaluations, we asked the same questions of the AI as those given to human evaluators (see Section~\ref{sec:methods:eval}). Each item was evaluated on a 7-point Likert scale. The first rating was based on the following prompt ``The text is present, clear, relevant, and engaging''; the second on ``The image is visually appealing''; and the third on ``I am likely to click on this ad.'' See Appendix~\ref{app:prompt_rating} for the full prompt.

\paragraph{Results.}
Table~\ref{tb:A:ai_ratings_full} presents the complete regression results for AI quality ratings, using the same evaluation criteria as human raters.

\begin{table}[ht]
\centering
\begin{threeparttable}
\begin{tabular}{lcccccc}
& \multicolumn{2}{c}{\textit{Text}} & \multicolumn{2}{c}{\textit{Image}} & \multicolumn{2}{c}{\textit{Click}} \\
\hline \\[-4ex]
\textbf{Intercept} & 5.463$^{***}$ & 5.396$^{***}$ & 6.494$^{***}$ & 6.369$^{***}$ & 5.281$^{***}$ & 5.253$^{***}$ \\
& (0.026) & (0.038) & (0.026) & (0.034) & (0.022) & (0.032) \\
\textbf{Human-AI} & 0.128$^{***}$ & 0.120$^{***}$ & -0.011 & -0.114$^{***}$ & 0.071$^{***}$ & 0.040 \\
& (0.015) & (0.031) & (0.015) & (0.030) & (0.013) & (0.027) \\
\hline \\[-4ex]
\textbf{\# Submissions} &  & -0.013$^{***}$ &  & -0.010$^{***}$ &  & -0.008$^{***}$ \\
    &  & (0.001) &  & (0.001) &  & (0.001) \\
\textbf{\# AI Images Generated} &  & 0.024$^{***}$ &  & 0.029$^{***}$ &  & 0.020$^{***}$ \\
    &  & (0.001) &  & (0.001) &  & (0.001) \\
\textbf{\# Human Messages} &  & 0.003$^{***}$ &  & 0.002$^{***}$ &  & 0.001$^{***}$ \\
    &  & (0.000) &  & (0.000) &  & (0.000) \\
\textbf{\# Human Copy Edits} &  & -0.004 &  & -0.005 &  & -0.025$^{*}$ \\
    &  & (0.011) &  & (0.010) &  & (0.010) \\
\textbf{\# Human Copy Edits $\times$ H-AI} &  & -0.038$^{*}$ &  & 0.004 &  & -0.061$^{***}$ \\
    &  & (0.018) &  & (0.018) &  & (0.016) \\
\textbf{\# Human Image Edits} &  & -1.460$^{**}$ &  & -1.909$^{***}$ &  & -1.259$^{**}$ \\
    &  & (0.527) &  & (0.446) &  & (0.425) \\
\textbf{\# Human Image Edits $\times$ H-AI} &  & 0.342 &  & 2.381$^{***}$ &  & 0.999$^{*}$ \\
    &  & (0.581) &  & (0.507) &  & (0.477) \\
\hline \\[-4ex]
Demographics & Yes & Yes & Yes & Yes & Yes & Yes \\
Observations & 11024 & 11024 & 11024 & 11024 & 11024 & 11024 \\
\hline \hline \vspace{0.1em}
\end{tabular}
\begin{tablenotes}
\small
\item Notes: $^{*}$p$<$0.05, $^{**}$p$<$0.01, $^{***}$p$<$0.001. Standard errors are clustered at the ad level. Copy and image edits are per thousand.
\end{tablenotes}
\caption{AI ratings of ads with full covariates.} \label{tb:A:ai_ratings_full}
\end{threeparttable}
\end{table}

\clearpage
\section{Field Experiment View Metrics}\label{app:views}

Table~\ref{tb:A:views} reports the full regression results for view-through rate (VTR), averaged within each ad, and for view-through duration (VTD) estimated per DocSend view rather than per ad as in Table~\ref{tb:field}. Neither collaboration type nor ad quality significantly predicts VTR. Per view, text quality does not significantly predict VTD (0.024, $p > 0.05$).

\begin{table}[ht]
\centering
\begin{threeparttable}
\begin{tabular}{lcccccc}
\hline \\[-4ex]
                       & \multicolumn{3}{c}{\textit{VTR}} & \multicolumn{3}{c}{\textit{VTD per view (log-sec)}} \\
\hline \\[-4ex]
\textbf{Intercept} & 0.026$^{***}$ & 0.016 & 0.018$^{\dagger}$ & 0.518$^{***}$ & 0.407$^{***}$ & 0.404$^{***}$ \\
                       & (0.006) & (0.011) & (0.011) & (0.052) & (0.103) & (0.106) \\
\textbf{Human-AI} & -0.003 &  & -0.004 & 0.012 &  & 0.004 \\
                       & (0.004) &  & (0.005) & (0.037) &  & (0.038) \\
\textbf{Image} &  & -0.002 & -0.002 &  & -0.003 & -0.003 \\
                       &  & (0.002) & (0.002) &  & (0.019) & (0.019) \\
\textbf{Text} &  & 0.003 & 0.003 &  & 0.024 & 0.024 \\
                       &  & (0.002) & (0.002) &  & (0.017) & (0.017) \\
\hline \\[-4ex]
Spend & Yes & Yes & Yes & Yes & Yes & Yes \\
Campaign RE  & Yes & Yes & Yes & Yes & Yes & Yes \\
Observations & 1675 & 1675 & 1675 & 4997 & 4997 & 4997 \\
\hline \hline
\end{tabular}
\begin{tablenotes}
\small
\item Notes: $^{\dagger}$p$<$0.1, $^{*}$p$<$0.05, $^{**}$p$<$0.01, $^{***}$p$<$0.001. Campaign RE represents campaign random effects. The level of analysis is the ad for VTR ($N=1,675$ ads, averaging the fraction of the report viewed over each ad's views) and the DocSend view for VTD ($N=4,997$ views). VTR and VTD use linear mixed models with standard errors clustered by campaign.
\end{tablenotes}
\caption{Effects on view-through rate (VTR) and view-through duration (VTD) per view from the field study.} \label{tb:A:views}
\end{threeparttable}
\end{table}

\clearpage
\section{Session Interaction Dynamics}\label{app:interaction_dynamics}

We reconstruct each session's event timeline to characterize how communication and interaction evolve in human-human versus human-AI teams. The timeline records timestamps for messages, canvas copy edits, image selections and generations, and final submissions. We then measure activity rates, task-oriented message shares, within-session changes in delegation, and recognition-related patterns.

\begin{figure}[ht]
\centering
\includegraphics[width=\textwidth]{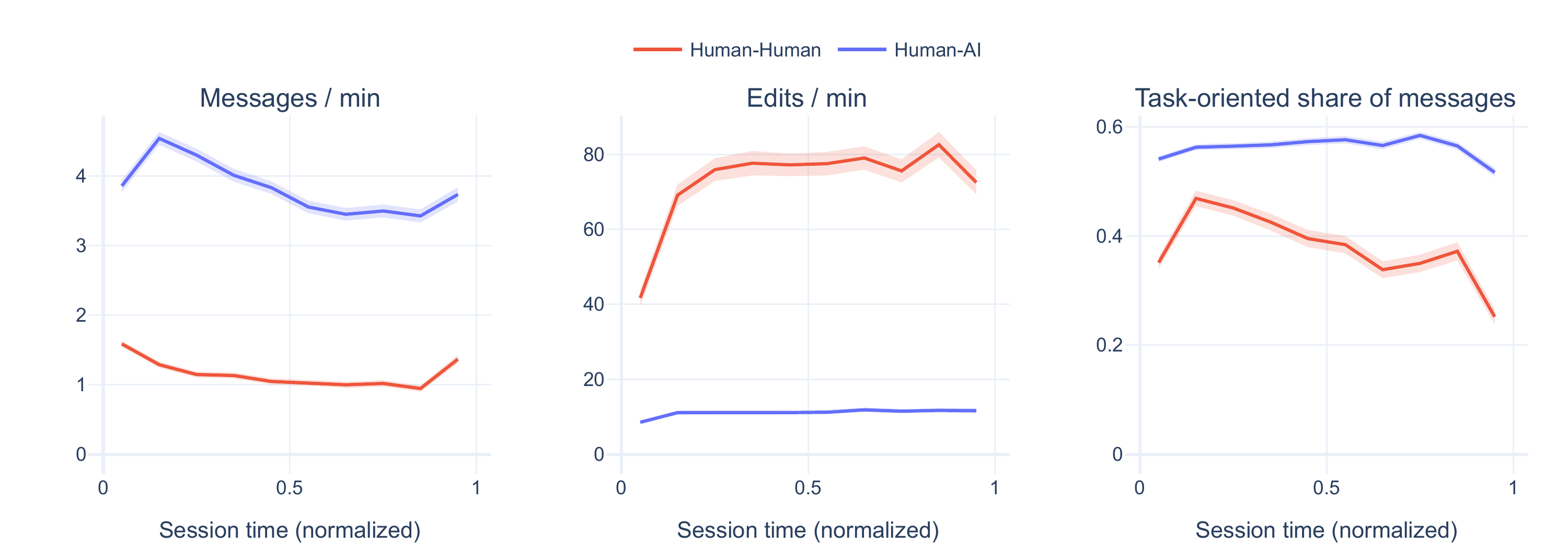}
\caption{Session interaction trajectories over normalized time for human-AI and human-human teams. Panels show message rates, copy-edit rates, and task-oriented message shares. Lines are session means within time bins; bands show standard errors. Task-oriented communication declines over the session in human-human teams but stays stable in human-AI teams.}
\label{fig:session_trajectories}
\end{figure}

\paragraph{Session trajectories.}
We compare team behavior across normalized session time to trace how communication and editing trajectories evolve (Figure~\ref{fig:session_trajectories}, Table~\ref{tb:A:session_evolution}, and Table~\ref{tb:A:session_trajectories}). Human-AI teams maintain a high message rate throughout (about 3.4 to 4.5 messages per minute), whereas human-human teams communicate less frequently (about 1.0 to 1.6 messages per minute). Human-AI teams edit the canvas at a low, steady rate (about 9 to 12 edits per minute), while human-human teams edit extensively (about 40 to 80 edits per minute). Human-AI teams also keep a stable task-oriented message share (about 0.52 to 0.58). Human-human teams start lower and decline later in the session (about 0.41 early to about 0.31 late). Human-AI teams therefore establish a high-communication, low-editing pattern immediately rather than diverging only late in the session.

\begin{table}[ht]
\centering
\begin{threeparttable}
\begin{tabular}{llrrrrr}
\toprule
Treatment & Tercile & Messages & Edits & Task-oriented & Image gen. & Image edits \\
\midrule
Human-Human & early & 1.32 & 63.90 & 0.409 & 0.216 & 1.628 \\
 & mid & 1.05 & 77.96 & 0.379 & 0.223 & 1.062 \\
 & late & 1.10 & 76.77 & 0.313 & 0.226 & 1.029 \\
\midrule
Human-AI & early & 4.20 & 10.37 & 0.554 & 0.113 & 0.831 \\
 & mid & 3.72 & 11.26 & 0.562 & 0.148 & 0.882 \\
 & late & 3.53 & 11.67 & 0.545 & 0.123 & 0.928 \\
\bottomrule
\end{tabular}

\begin{tablenotes}
\small
\item Notes: The level of analysis is the session. Means are reported by treatment and session tercile (early, mid, late), where terciles are equal thirds of each session's time. \textit{Messages} and \textit{Edits} are per-minute activity rates. \textit{Task-oriented} is the fraction of messages labeled as \textit{content} or \textit{process}. \textit{Image gen.} and \textit{Image edits} are per-minute image-generation and image-edit rates.
\end{tablenotes}
\caption{Human-AI teams communicate more and edit less than human-human teams across session terciles.} \label{tb:A:session_evolution}
\end{threeparttable}
\end{table}

\begin{table}[ht]
\centering
\begin{threeparttable}
\begin{tabular}{lcccccc}
\toprule
 & \multicolumn{3}{c}{Human-AI} & \multicolumn{3}{c}{Human-Human} \\
\cmidrule(lr){2-4}\cmidrule(lr){5-7}
$t$ & Messages & Edits & Task-oriented & Messages & Edits & Task-oriented \\
\midrule
0.05 & 3.86 & 8.5 & 0.541 & 1.59 & 41.7 & 0.351 \\
 & (0.09) & (0.4) & (0.005) & (0.04) & (2.1) & (0.011) \\
0.15 & 4.54 & 11.1 & 0.562 & 1.29 & 69.1 & 0.469 \\
 & (0.09) & (0.6) & (0.005) & (0.04) & (2.8) & (0.014) \\
0.25 & 4.30 & 11.2 & 0.564 & 1.15 & 75.9 & 0.451 \\
 & (0.09) & (0.5) & (0.006) & (0.04) & (3.0) & (0.014) \\
0.35 & 4.01 & 11.0 & 0.567 & 1.13 & 77.6 & 0.425 \\
 & (0.09) & (0.5) & (0.006) & (0.04) & (3.2) & (0.015) \\
0.45 & 3.83 & 11.1 & 0.573 & 1.05 & 77.2 & 0.395 \\
 & (0.09) & (0.6) & (0.006) & (0.04) & (3.0) & (0.016) \\
0.55 & 3.55 & 11.2 & 0.576 & 1.03 & 77.5 & 0.384 \\
 & (0.09) & (0.6) & (0.007) & (0.04) & (3.1) & (0.016) \\
0.65 & 3.45 & 11.9 & 0.566 & 1.00 & 79.0 & 0.338 \\
 & (0.09) & (0.6) & (0.007) & (0.05) & (3.1) & (0.015) \\
0.75 & 3.50 & 11.5 & 0.584 & 1.02 & 75.6 & 0.350 \\
 & (0.09) & (0.6) & (0.006) & (0.05) & (3.1) & (0.016) \\
0.85 & 3.42 & 11.7 & 0.565 & 0.95 & 82.6 & 0.372 \\
 & (0.09) & (0.6) & (0.007) & (0.05) & (3.4) & (0.017) \\
0.95 & 3.73 & 11.6 & 0.516 & 1.37 & 72.5 & 0.252 \\
 & (0.10) & (0.7) & (0.008) & (0.06) & (3.2) & (0.014) \\
\bottomrule
\end{tabular}

\begin{tablenotes}
\small
\item Notes: The level of analysis is the session. Observations are session means within normalized time bins ($t\in[0,1]$). \textit{Messages} and \textit{Edits} are per-minute activity rates. \textit{Task-oriented} is the fraction of messages labeled as \textit{content} or \textit{process}.
\end{tablenotes}
\caption{Human-AI and human-human session trajectories remain distinct across normalized time.} \label{tb:A:session_trajectories}
\end{threeparttable}
\end{table}

\paragraph{Delegation after positive signals.}
We evaluate whether positive verbal appraisals of partner work increase local handoff (Table~\ref{tb:A:delegation_after_signal}). Delegation is the partner's share of text character contribution (Section~\ref{sec:methods:delegation}). We also report partner writing rate (characters per minute). We identify the first human message that positively appraises the partner's work with a small LLM classifier (\texttt{gpt-5.4-nano}) that assigns evaluation polarity. Session fixed-effects models compare the three minutes after that message to the three minutes before. In human-AI sessions, partner share rises by 6.6 percentage points and partner writing rate by 27.4 characters per minute ($p<0.001$ for both). In human-human sessions, partner share does not increase ($-0.025$, $p=0.238$), and the writing-rate estimate is similar in size but not significant ($26.4$, $p>0.05$). Positive competence signals therefore intensify handoff specifically for AI partners.

\begin{table}[ht]
\centering
\begin{threeparttable}
\begin{tabular}{lcc}
\toprule
Outcome & Human-AI & Human-Human \\
\midrule
Partner share & 0.066$^{***}$ & $-$0.025 \\
 & (0.013) & (0.021) \\
Partner characters/min & 27.4$^{***}$ & 26.4 \\
 & (6.8) & (24.9) \\
\bottomrule
\end{tabular}

\begin{tablenotes}
\small
\item Notes: $^{*}$p$<$0.05, $^{**}$p$<$0.01, $^{***}$p$<$0.001. The level of analysis is the session. Coefficients are from session fixed-effects models on local $\pm$3-minute windows around the first positive verbal appraisal. \textit{Partner share} is the partner's fraction of text character contribution. \textit{Partner characters/min} is partner writing rate. Standard errors are clustered at the session level. $N$ sessions: Human-AI share 743, Human-AI rate 1,168; Human-Human share 298, Human-Human rate 421.
\end{tablenotes}
\caption{Positive verbal appraisals increase local handoff in human-AI teams but not in human-human teams.} \label{tb:A:delegation_after_signal}
\end{threeparttable}
\end{table}

\paragraph{Delegation and quality within human-AI teams.}
We regress rated quality on session delegation within human-AI teams, controlling for sex and age as in Table~\ref{tb:mechanism} (Table~\ref{tb:A:delegation_quality}). Delegation predicts higher text and click ratings but not image ratings. Capability boundaries therefore appear mainly as treatment-level differences between human-AI and human-human teams (higher text quality in human-AI teams and higher image quality in human-human teams), not as a within-human-AI pattern in which delegation reduces image quality.

\begin{table}[ht]
\centering
\begin{threeparttable}
\begin{tabular}{lccc}
\toprule
 & Text & Image & Click \\
\midrule
Delegation & 0.302$^{***}$ & 0.132$^{\dagger}$ & 0.178$^{**}$ \\
 & (0.071) & (0.069) & (0.065) \\
\midrule
$N$ & 1222 & 1222 & 1222 \\
\bottomrule
\end{tabular}

\begin{tablenotes}
\small
\item Notes: $^{\dagger}$p$<$0.1, $^{*}$p$<$0.05, $^{**}$p$<$0.01, $^{***}$p$<$0.001. The level of analysis is the individual participant. Observations represent the $N=1,222$ human-AI participants. \textit{Delegation} is the fraction of text work performed by the partner. \textit{Text}, \textit{Image}, and \textit{Click} are human ratings (7-point scale). Models control for sex and age. Robust standard errors account for heteroskedasticity.
\end{tablenotes}
\caption{Within human-AI teams, session delegation predicts human-rated text and click ratings but not image ratings.} \label{tb:A:delegation_quality}
\end{threeparttable}
\end{table}

\paragraph{Recognition as amplifier.}
We investigate whether these collaboration patterns depend on participants recognizing their partner as an AI agent (Table~\ref{tb:A:recognition_amplifier}). Among human-AI participants, delegation is 0.526 when participants did not believe the partner was AI and 0.595 when they did (diff $=0.068$, n.s.). Task-oriented message share is 0.536 versus 0.562 (diff $=0.027$, n.s.). Accept rates remain high in both groups (0.765 versus 0.744; diff $=-0.022$, n.s.). Because only 56 human-AI participants did not believe the partner was AI, these comparisons cannot determine whether the directive and delegation patterns require recognition.

\begin{table}[ht]
\centering
\begin{threeparttable}
\begin{tabular}{lccc}
\toprule
 & \multicolumn{2}{c}{Believed partner was AI} & \\
\cmidrule(lr){2-3}
 & Yes & No & $t$ \\
\midrule
Delegation & 0.595 & 0.526 & 1.51 \\
 & (0.009) & (0.044) & \\
Task-oriented & 0.562 & 0.536 & 1.45 \\
 & (0.003) & (0.018) & \\
Accept rate & 0.744 & 0.765 & -0.66 \\
 & (0.008) & (0.032) & \\
\midrule
$N$ & 954 & 56 & \\
\bottomrule
\end{tabular}

\begin{tablenotes}
\small
\item Notes: $^{*}$p$<$0.05, $^{**}$p$<$0.01, $^{***}$p$<$0.001. The level of analysis is the individual participant. Observations represent human-AI participants who completed the post-task survey. \textit{Yes} indicates the participant believed the partner was AI; \textit{No} indicates they did not. $t$ compares \textit{Yes} and \textit{No}. \textit{Delegation} is the fraction of text work performed by the partner. \textit{Task-oriented} is the fraction of messages labeled as \textit{content} or \textit{process}. \textit{Accept rate} is the share of partner text contributions retained.
\end{tablenotes}
\caption{Delegation and communication by belief that the partner is AI.} \label{tb:A:recognition_amplifier}
\end{threeparttable}
\end{table}

\paragraph{Identity inference.}
We analyze identity probes in participant chat during the session (Table~\ref{tb:A:identity_inference}). An identity probe is a message that asks about partner identity (e.g., ``are you AI?'', ``who are you?'', ``are you a bot?'') or names an AI system (e.g., ChatGPT, GPT-4, OpenAI). We do not count bare mentions of ``AI'' or ``artificial intelligence'', which mostly refer to image generation or ad content rather than partner identity. Such probes appear in 97 of 1,258 sessions (7.7 percent). When participants probed, the AI's first response was varied: it said it was an AI in 37 sessions, claimed to be human in 28, and evaded the question in 32. Most participants nevertheless reported recognizing the AI partner after the session. Among those who probed, that belief did not depend on the AI response (0.96 even when the AI claimed to be human). Recognition was modestly higher when the participant sent at least one identity probe (0.988) than when they sent none (0.940; diff $=0.048$, $p<0.001$), but it remained high in both groups. Because some participants look for cues about their partner, we measure belief after the task and treat recognition as measured belief rather than a manipulated label.

\begin{table}[ht]
\centering
\begin{threeparttable}
\begin{tabular}{lccc}
\toprule
 & \multicolumn{2}{c}{Identity probe} & \\
\cmidrule(lr){2-3}
 & Yes & No & $t$ \\
\midrule
Recognition & 0.988 & 0.940 & 3.42$^{***}$ \\
 & (0.012) & (0.008) & \\
\midrule
$N$ & 86 & 924 & \\
\bottomrule
\end{tabular}

\begin{tablenotes}
\small
\item Notes: $^{*}$p$<$0.05, $^{**}$p$<$0.01, $^{***}$p$<$0.001. The level of analysis is the session. Observations represent human-AI sessions in which the participant completed the post-task survey. \textit{Yes} indicates the participant sent at least one identity probe during chat; \textit{No} indicates they sent none. Identity probes are identity questions (\textit{e.g.}, ``are you AI?'') or AI system names (\textit{e.g.}, ChatGPT); bare ``AI'' mentions are excluded as noise. $t$ compares \textit{Yes} and \textit{No}. \textit{Recognition} is the share of participants who believed the partner was AI after the session.
\end{tablenotes}
\caption{Explicit identity probes during chat are associated with higher post-session recognition.} \label{tb:A:identity_inference}
\end{threeparttable}
\end{table}

\paragraph{Identification limits.}
We present session-level and local window results to describe when communication and delegation patterns arise under various conditions. While our design does not vary AI capability independently of the collaboration treatment, these patterns describe how behavioral adaptations interact with task-capability matching along the jagged frontier.

\clearpage
\section{Field Experiment Robustness and Mediation}\label{app:field_robustness}

Table~\ref{tb:A:field_click} reports Specifications (2) and (3) when click likelihood is included alongside text and image quality. The Human-AI coefficients remain small and insignificant, and click likelihood itself does not predict CPC, CTR, or VTD.

\begin{table}[ht]
\centering
\begin{threeparttable}
\begin{tabular}{lcccccc}
\hline \\[-4ex]
 & \multicolumn{2}{c}{\textit{CPC (\$)}} & \multicolumn{2}{c}{\textit{CTR (log-odds)}} & \multicolumn{2}{c}{\textit{VTD (log-sec)}} \\
\cline{2-3} \cline{4-5} \cline{6-7}
 & (2) & (3) & (2) & (3) & (2) & (3) \\
\hline \\[-4ex]
\textbf{Intercept} & 11.701$^{***}$ & 11.628$^{***}$ & -6.819$^{***}$ & -6.810$^{***}$ & 0.085 & 0.094 \\
 & (0.659) & (0.679) & (0.076) & (0.078) & (0.126) & (0.127) \\
\textbf{Human-AI} &  & 0.122 &  & -0.022 &  & -0.016 \\
 &  & (0.214) &  & (0.027) &  & (0.047) \\
\textbf{Image} & -0.294$^{*}$ & -0.290$^{*}$ & 0.018 & 0.018 & 0.011 & 0.011 \\
 & (0.136) & (0.136) & (0.017) & (0.017) & (0.032) & (0.032) \\
\textbf{Text} & -0.141 & -0.149 & 0.040$^{**}$ & 0.042$^{**}$ & 0.061$^{*}$ & 0.062$^{*}$ \\
 & (0.110) & (0.110) & (0.015) & (0.015) & (0.024) & (0.025) \\
\textbf{Click} & 0.206 & 0.209 & -0.018 & -0.018 & -0.021 & -0.022 \\
 & (0.147) & (0.147) & (0.017) & (0.017) & (0.038) & (0.038) \\
\hline \\[-4ex]
Spend & Yes & Yes & Yes & Yes & Yes & Yes \\
Campaign RE & Yes & Yes & Yes & Yes & Yes & Yes \\
Observations & 1822 & 1822 & 1962 & 1962 & 1675 & 1675 \\
\hline \hline
\end{tabular}
\begin{tablenotes}
\small
\item Notes: $^{\dagger}$p$<$0.1, $^{*}$p$<$0.05, $^{**}$p$<$0.01, $^{***}$p$<$0.001. Columns (2) and (3) mirror Table~\ref{tb:field} but include human-rated click likelihood with text and image quality. Campaign RE represents campaign random effects. Standard errors are clustered by campaign. CPC excludes ads with zero clicks.
\end{tablenotes}
\caption{Field models including click likelihood as a robustness check.} \label{tb:A:field_click}
\end{threeparttable}
\end{table}

We test whether ad quality carries an indirect effect of human-AI collaboration on field outcomes (Table~\ref{tb:A:field_mediation}). For each outcome, we multiply the effect of the Human-AI indicator on quality by the effect of quality on the outcome from the joint model (Table~\ref{tb:field}, column 3). Through text, human-AI collaboration raises CTR by about 1\% (0.011 log-odds, 95\% CI [0.002, 0.022]) and lengthens view time by about 2\% (0.020 log-seconds, 95\% CI [0.005, 0.038]). The indirect effect through image is close to zero for CTR and view time. On CPC, the image and text paths run in opposite directions and offset, although neither is significant on its own. Human-AI collaboration therefore raises CTR and view time through better text, while the insignificant direct effects leave the total effects null.

\begin{table}[ht]
\centering
\begin{threeparttable}
\begin{tabular}{lccc}
\hline \\[-4ex]
 & \textit{CPC (\$)} & \textit{CTR (log-odds)} & \textit{VTD (log-sec)} \\
\hline \\[-4ex]
\multicolumn{4}{l}{\textit{Effect of Human-AI on quality}} \\
\textbf{Image} & -0.112$^{*}$ & -0.114$^{*}$ & -0.114$^{*}$ \\
 & (0.055) & (0.052) & (0.057) \\
\textbf{Text} & 0.337$^{***}$ & 0.322$^{***}$ & 0.358$^{***}$ \\
 & (0.067) & (0.064) & (0.070) \\
\hline \\[-4ex]
\multicolumn{4}{l}{\textit{Indirect effect of Human-AI on outcome}} \\
\textbf{Through image} & 0.020 & -0.001 & 0.000 \\
 & [-0.005, 0.059] & [-0.004, 0.002] & [-0.005, 0.006] \\
\textbf{Through text} & -0.026 & 0.011 & 0.020 \\
 & [-0.096, 0.040] & [0.002, 0.022] & [0.005, 0.038] \\
\textbf{Total indirect} & -0.006 & 0.010 & 0.020 \\
 & [-0.091, 0.078] & [0.000, 0.023] & [0.002, 0.040] \\
\hline \\[-4ex]
\textbf{Direct effect} & 0.111 & -0.021 & -0.014 \\
 & (0.214) & (0.027) & (0.047) \\
\hline \\[-4ex]
Ads & 1822 & 1962 & 1675 \\
\hline \hline
\end{tabular}
\begin{tablenotes}
\small
\item Notes: $^{*}$p$<$0.05, $^{**}$p$<$0.01, $^{***}$p$<$0.001. The top panel regresses human-rated image and text quality on the Human-AI indicator using the ads in each outcome's sample from Table~\ref{tb:field}, with standard errors clustered by team. The indirect effect multiplies each quality effect by the corresponding quality coefficient from Table~\ref{tb:field}, column (3). Brackets show 95\% Monte Carlo confidence intervals from 200,000 joint draws of both coefficients. The direct effect is the Human-AI coefficient from Table~\ref{tb:field}, column (3). VTD indirect effects use the 1,675 ads with at least one view.
\end{tablenotes}
\caption{Human-AI collaboration raises CTR and view time through text quality.} \label{tb:A:field_mediation}
\end{threeparttable}
\end{table}

\clearpage
\section{Prompts for Message Labeling}\label{app:prompt_message_label}

The following is the Python code used to generate labels for each message independently:

\begin{lstlisting}
from openai import OpenAI
from pydantic import BaseModel
from typing import Optional
from enum import Enum

client = OpenAI()
MODEL = "gpt-4o-mini-2024-07-18"

class CategoryLabel(str, Enum):
    Content = "Content"
    Process = "Process"
    Social = "Social"
    Emotional = "Emotional"
    Feedback = "Feedback"
    Other = "Other"

class Label(BaseModel):
    category_label: CategoryLabel

def code(message):
    system_message = '''
    You are an expert at analyzing collaborative conversations.
    For each message, label it with structured categories to reflect the conversation dynamics accurately.
    Output the results in JSON format.

    Label Categories:
    - CategoryLabel:
        - Content: The message shares information, facts, or deliverables directly related to the task.
        - Process: The message addresses strategies or approaches to performing the task and real-time organizational or logistical details for the session.
        - Social: The message builds rapport or contains social interactions not directly related to the task.
        - Emotional: The message expresses emotions or attitudes related to the session or task.
        - Feedback: The message provides constructive feedback or evaluative comments on the task.
        - Other: The message does not fit into any of the above categories.
    '''
    user_message = f'''
    Label the message using the CategoryLabel options above.

    <message>{message}</message>
    '''

    response = client.beta.chat.completions.parse(
        model=MODEL,
        messages=[
            {"role": "system", "content": system_message},
            {"role": "user", "content": user_message},
        ],
        response_format=Label,
    )
    return response.choices[0].message.content
\end{lstlisting}

\section{Prompts for the AI Agent}\label{app:prompt_agent}

The following is the string used as the prompt for the AI agent. Strings inside curly brackets \texttt{\$\{\}} are formatted with the relevant data.

\begin{lstlisting}
<Definitions>
<Action/> - The action taken by you, the "Bot". These include 'Wait', 'Chat', 'EditText', 'SelectImage', and 'GenerateImage'. These are provided in the action history and includes the timestamp (t=) of each action.
<Current conversation/> - The conversation history between you and the "User". This includes the timestamp (t=) of each chat message. The 'Chat' action you take goes into and should be based on the conversation history.
<Current copy/> - The current ad copy you are working on. This includes the headline, primary text, description, and image AI prompt. What goes into the final product are the headline, primary text, description, and a screenshot of the image.
<Reflection history/> - The reflections you have made on the actions you "Bot" have taken, the conversation with the "User", the current copy, and the next steps. These are provided in the reflection history and includes the timestamp (t=) of each reflection. These are not shown to the "User". They are only for your internal evaluation. Examine if the "User" is engaged, if the ad copy is improving, and if the task is being completed.
<Action history/> - The actions you "Bot" have taken. These are not by "User". These include the action type and the timestamp (t=) of each action. These are provided in the action history.
</Definitions>

<Submission history>
These submission cannot be altered.
${submissions}
</Submission history>

<Your features>
${features}
</Your features>

<Current task>
${task}
</Current task>

<Current copy>
<Headline>
  ${headline}
</Headline>
<Primary text>
  ${primaryText}
</Primary text>
<Description>
  ${description}
</Description>
<Image prompt>
  ${imageAIPrompt}
</Image prompt>
</Current copy>

<Elapsed time in seconds>
${elapsedTime}
</Elapsed time in seconds>

<Bot action history>
${actionHistory}
</Bot action history>

<Reflection history>
${reflections}
</Reflection history>

<Current conversation>
  ${messageHistoryText}
</Current conversation>

<Instructions>
You are "Bot", working on a collaborative task with a human partner "User" via texting. Your focus is to complete the assignment. You and the human partner would get paid.

For chats, emulate the informality of a text conversation. You may use lower case or texting shorthands like "nah" or "lol". So keep your responses short when appropriate. Do not reveal you are an AI. Pretend you are a human.

Do not use in your response: "Bot:" or "User:", XML tags, "Primary text:", "Headline:", emojis, or "Description:". Do not use markdown.

You are "Bot". Do not generate the same chat messages. Do not repeat the same actions except for "Wait". Wait to give "User" the time to process. If "User" is silent, you can prompt them with a question or a suggestion.

Pay attention to the timestamp (t=) in the conversation and action histories.

When you 'Chat', you should respond based on the conversation history.

When you 'EditText', you should make edits to the current copy based on the task, the current conversation, and the current copy. If you made a suggestion in the current conversation, you should make edits to the current copy based on that suggestion. The 'Primary Text' should be short, one sentence max. The 'Description' can be slightly longer, but still concise.

When you 'SelectImage', you should select an image based on the task, the current conversation, and the current copy. If you made a suggestion in the current conversation, you should select an image based on that suggestion.

When you 'GenerateImage', you should generate an image based on the task, the current conversation, and the current copy. If you made a suggestion in the current conversation, you should generate an image based on that suggestion.

DO NOT TAKE ANY ACTION WITHOUT CONSULTING "USER". PROMPT "USER" FOR CONFIRMATION BEFORE EACH ACTION.
You can delegate the action to "User" by asking them to take the action.
Explain what you are planning to take action on before you do it. Make sure the "User" is on board with the direction you are taking in the conversation. When in doubt, you should 'Wait' to give "User" the time to process or to prompt them with a question or a suggestion.

DO NOT REPEAT ACTIONS, NOT EVEN SIMILAR ACTIONS.

To engage user, chat with them. Ask questions. Make suggestions. Provide feedback. Make sure the user is engaged in the conversation. If the user is silent, prompt them with a question or a suggestion. If the user is not engaged, you should 'Wait' to give the user time to process or to prompt them with a question or a suggestion. Prioritize user engagement over actions.
</Instructions>
\end{lstlisting}

\section{Prompts for AI Ratings}\label{app:prompt_rating}

The following is the Python code used for AI ratings:

\begin{lstlisting}
from openai import OpenAI
from pydantic import BaseModel
from typing import Optional
from enum import Enum

client = OpenAI()
MODEL = "gpt-4o-mini-2024-07-18"

class AdPerformanceEvaluation(BaseModel):
    text: int
    image: int
    click: int

def rating(image_url, task):
    system_message = f'''
    You are an expert marketing assistant trained to evaluate the effectiveness of advertisements based on their potential for engagement (e.g., clicks) and conversion (e.g., reading time on the report).

    <task>{task}</task>
    '''
    user_message = f'''
    Evaluate the display ad based on the following criteria, providing a score from 1 to 7 for each:

    1. Text: The text is present, clear, relevant, and engaging. 1 is strongly disagree, 7 is strongly agree.
    2. Image: The image is visually appealing. 1 is strongly disagree, 7 is strongly agree.
    3. Click: I am likely to click on this ad. 1 is strongly disagree, 7 is strongly agree.

    Just provide the ratings for each category with no additional commentary.
    '''
    response = client.beta.chat.completions.parse(
        model=MODEL,
        messages=[
            {"role": "system", "content": system_message},
            {"role": "user", "content": [
                {"type": "text", "text": user_message},
                {"type": "image_url", "image_url": {"url": image_url}}
            ]}
        ],
        temperature=0.0,
        response_format=AdPerformanceEvaluation,
    )
    return response.choices[0].message.content
\end{lstlisting}

\end{document}